\documentclass[a4paper,fleqn]{cas-sc}

\usepackage{subfiles}

\usepackage[super,sort&compress]{natbib}

\usepackage[utf8]{inputenc} 
\usepackage[T1]{fontenc} 
\usepackage{array} 
\usepackage{tabularx} 
\usepackage{float} 
\usepackage{multirow} 
\usepackage{threeparttable} 
\usepackage[section]{placeins} 
\usepackage{textgreek} 
\usepackage{url} 
\graphicspath{{./}{figures/}} 

\usepackage{xcolor}
\usepackage{caption}
\makeatletter
\@ifundefined{orcidname}{}{}
\@ifundefined{printorcid}{}{}
\makeatother

\def\tsc#1{\csdef{#1}{\textsc{\lowercase{#1}}\xspace}}
\tsc{WGM}
\tsc{QE}
\tsc{EP}
\tsc{PMS}
\tsc{BEC}
\tsc{DE}

\begin{document}

\let\WriteBookmarks\relax
\def\floatpagepagefraction{1}
\def\textpagefraction{.001}
\shorttitle{Accelerating gas-network feasibility screening}
\shortauthors{D. Jiang et~al.}


\title [mode = title]{Accelerating gas-network feasibility screening with a physics-informed graph neural network surrogate}
%
%

\author[1,2]{Dongrui Jiang}\cormark[1]
\ead{dongrui.jiang@scai.fraunhofer.de}
\credit{Conceptualization, Methodology, Software, Investigation, Formal analysis, Visualization, Writing -- original draft, Writing -- review \& editing}

\author[2,3]{Jochen Garcke}
\credit{Supervision, Methodology, Formal analysis, Writing -- review \& editing}

\author[1,2]{Okan Akca}
\credit{Investigation, Data curation, Validation, Writing -- review \& editing}

\author[1]{Jeremias Hollnagel}
\credit{Resources, Data curation, Writing -- review \& editing}

\author[4]{Bernhard Klaassen}
\credit{Resources, Writing -- review \& editing}

\author[2]{Mehrnaz Anvari}
\credit{Resources, Funding acquisition, Writing -- review \& editing}

\author[1]{Joachim M\"uller-Kirchenbauer}
\credit{Supervision, Resources, Writing -- review \& editing}

\affiliation[1]{organization={Technische Universit\"at Berlin},
                addressline={Stra\ss e des 17. Juni 135},
                postcode={10623},
                city={Berlin},
                country={Germany}}

\affiliation[2]{organization={Fraunhofer Institute for Algorithms and Scientific Computing SCAI},
                addressline={Schloss Birlinghoven 1},
                postcode={53757},
                city={Sankt Augustin},
                country={Germany}}

\affiliation[3]{organization={Institut für Numerische Simulation, Universit\"at Bonn},
                addressline={Friedrich-Hirzebruch-Allee 7},
                postcode={53115},
                city={Bonn},
                country={Germany}}

\affiliation[4]{organization={Fraunhofer Research Institution for Energy Infrastructures and Geothermal Systems IEG},
                addressline={Am Hochschulcampus 1},
                postcode={44801},
                city={Bochum},
                country={Germany}}

\cortext[cor1]{Corresponding author}


\begin{abstract}
Large-scale gas-network scenario evaluation is a computational bottleneck in integrated energy-system planning, particularly when gas infrastructure interacts with power, heat, hydrogen, and sector-coupling pathways. Conventional nonlinear hydraulic solvers provide reliable feasibility assessment but are costly for stochastic screening, whereas unconstrained learning-based surrogates may produce hydraulically infeasible states. This study develops a physics-informed graph neural network surrogate for steady-state gas-network simulation and feasibility screening. The model uses an edge-centric architecture to predict pipe-level squared-pressure differences and flows. A differentiable projection layer enforces nodal mass conservation on predicted flows, while a Laplacian reconstruction maps edge pressure differences to topologically consistent nodal pressures.

The framework is evaluated on GasLib-134, GasLib-135, and GasLib-582 using stochastically generated operating scenarios. On the meshed 582-node benchmark trained with 5000 scenarios, the surrogate achieves a pressure mean absolute error of 1.05~bar, corresponding to 1.3\% of the realized pressure range, with $R^2 = 0.981$. Projected-flow predictions reach $R^2 = 0.972$, and mass-balance residuals are reduced to numerical precision, on the order of $10^{-5}$--$10^{-4}$~Nm$^3$/s. Compared with the MYNTS reference solver, inference is reduced from seconds to milliseconds, with the largest benchmark evaluated in less than 40~ms. Loadability and out-of-distribution stress-test evaluations demonstrate robust feasibility screening under high-load conditions, while strongly localized demand concentrations are identified as cases requiring solver-based verification near feasibility limits. The framework provides a physically constrained planning accelerator for high-volume scenario screening and prioritization.
\end{abstract}

\begin{graphicalabstract}
\centering
\includegraphics[width=1.0\linewidth]{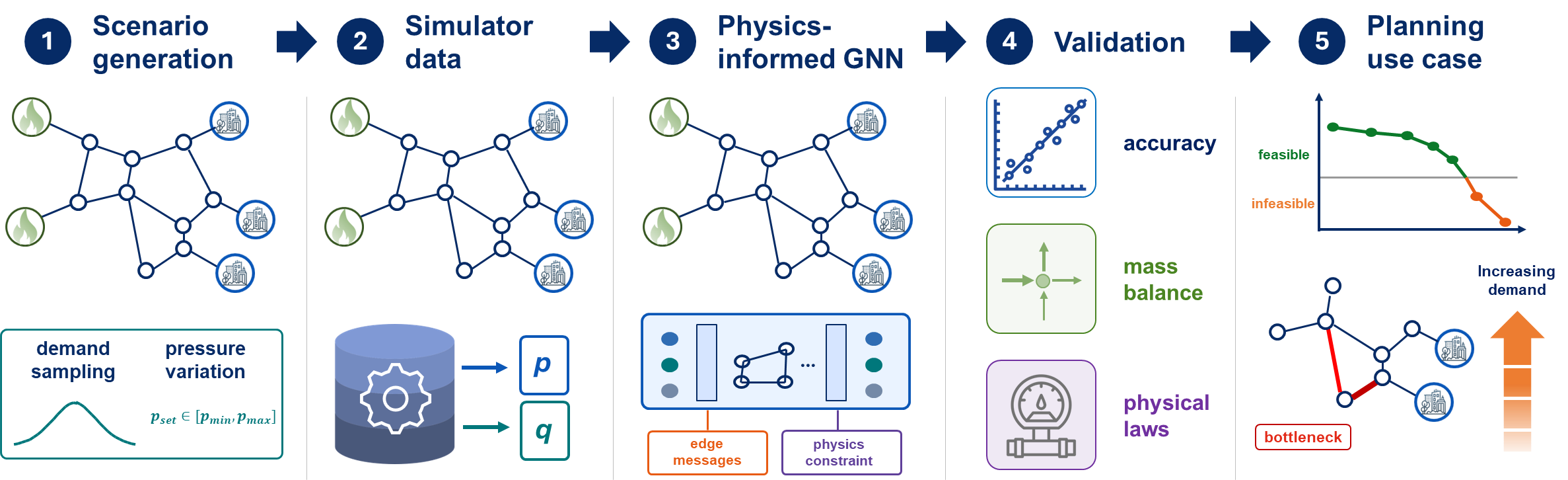}
\end{graphicalabstract}

\begin{highlights}
\item Edge-centric PI-GNN predicts pipe-level pressure drops and gas flows
\item Projection enforces mass balance; reconstruction improves pressure consistency
\item Achieves 1.05~bar pressure MAE and $R^2=0.972$ flow accuracy on GasLib-582
\item OOD stress tests identify cases needing solver verification
\item Enables millisecond-scale feasibility and bottleneck screening for planning
\end{highlights}


\maketitle


\section{Introduction}

Under deep decarbonization pathways, the long-term use of natural gas as a primary fossil fuel is expected to decline~\cite{IEA2025WEO,EnergyInstitute2025Review}. Nevertheless, existing gas transmission infrastructure is likely to remain relevant for future energy systems due to its large-scale transport capacity and storage flexibility, which are difficult to reproduce with electricity networks alone~\cite{Brown2018,Ramsebner2021}. In integrated multi-energy systems, these assets may support hydrogen transport, power-to-gas (P2G) conversion, localized gas injection, and bidirectional coupling with the power sector~\cite{Wang2023,Wietschel2020,Taibi2018,Mahajan2022}. The planning question is therefore not only whether gas infrastructure remains useful, but whether its hydraulic limits can be evaluated quickly across many uncertain operating conditions. This need builds on recent gas-network modeling efforts that embed spatially resolved simulation into larger energy-system workflows~\cite{Jiang2020,MuellerKirchenbauer2024}. The planning question thus creates a computational bottleneck: conventional hydraulic solvers are too slow for screening thousands of scenarios, while unconstrained surrogates risk physical infeasibility.

This requirement creates a computational bottleneck for planning studies. Uncertain demand and renewable generation~\cite{Chen2017,Li2023}, supply allocation and infrastructure changes~\cite{Ye2023,Clegg2016}, along with hydrogen blending and localized P2G injection~\cite{Kazi2024,HydrogenReview2024,Wang2025}, can generate large ensembles of operating scenarios that must be screened for pressure feasibility and potential bottlenecks. In such workflows, individual hydraulic simulations are less important than the ability to evaluate thousands of scenarios consistently and to identify the cases or network regions that require detailed engineering analysis. Efficient steady-state simulation is therefore a practical requirement for scenario-based stochastic planning, uncertainty analysis, and early-stage infrastructure assessment.

Conventional steady-state gas network simulation relies on nodal mass conservation and pipeline momentum balance, which are typically represented by nonlinear pressure--flow relations such as Weymouth-type equations~\cite{Osiadacz1987GasNetworks,Osiadacz1988SteadyState,Menon2005GasPipelineHydraulics}. These physics-based formulations provide sufficient fidelity for steady-state hydraulic design, planning, and feasibility studies, and are widely used in gas-network simulation and assessment tools~\cite{GanterEtAl2024}. However, these nonlinear systems are commonly solved using iterative methods such as Newton--Raphson, whose computational burden can increase with network size, nonlinear coupling, and repeated solution of the linearized systems arising within each iteration~\cite{BabonneauNesterovVial2012,LeongAyala2014}. Although such solvers are suitable for detailed single-case assessment, many-query analyses over large uncertainty or disruption scenario sets can become computationally prohibitive~\cite{HimpeGrundelBenner2021,GanterEtAl2024}. Because the solver output is defined implicitly by nonlinear residual equations rather than by a fixed explicit map, direct use in automatic-differentiation workflows can require implicit-differentiation or custom adjoint treatments for gradient-based sensitivity analysis and optimization~\cite{Margossian2019AD,BlondelEtAl2022ImplicitDiff}.

Surrogate and reduced-order modeling offer a route to high-throughput assessment in many-query gas and energy-network analyses~\cite{HimpeGrundelBenner2021}. Similar trends toward AI-assisted surrogate modeling and physics-informed computational acceleration are also emerging in hydrogen-energy technologies, including water-electrolyzer modeling, where machine-learning surrogates are increasingly used to complement computationally intensive multiphysics simulations~\cite{Alibeigi2026}. Planning applications impose stricter requirements than fast regression alone: a surrogate used for feasibility screening must return hydraulic states that are physically admissible enough to support decisions about pressure limits, flow redistribution, and limiting components in gas-transportation planning and optimization~\cite{RiosMercadoBorrazSanchez2015}. Recent work in physics-informed machine learning and graph-based modeling provides useful building blocks for this purpose. Physics-informed neural networks incorporate governing equations through residual or constraint losses, while graph neural networks exploit graph-structured physical systems through localized message passing~\cite{RaissiPerdikarisKarniadakis2019PINNs,Karniadakis2021PIML,Willard2022ACMSurvey,VonRueden2023InformedML,Wu2021GNNsurvey,KerimovEtAl2022WDSGNN}. Hybrid and constraint-aware approaches further attempt to combine data-driven flexibility with physical structure~\cite{Mohan2023HardConstraints,LuPestourie2021HardPINN}. GNN-based surrogates can accelerate simulation of meshed district-heating networks, yet they also reveal challenges in long-range coupling and receptive-field limitations for meshed energy networks~\cite{Yang2024}. In gas and energy networks, surrogate and reduced-order approaches have shown promise for accelerating many-query simulation and optimization, but maintaining strict hydraulic feasibility, robustness under off-design conditions, and scalability across meshed network structures remains challenging~\cite{RiosMercadoBorrazSanchez2015,HimpeGrundelBenner2021,GanterEtAl2024}.

The remaining gap is specific to planning-oriented gas-network surrogates. Purely data-driven models can achieve low pointwise prediction errors while still producing states that violate nodal mass conservation, pressure--flow consistency, or other hydraulic feasibility requirements. Physics-informed models based primarily on soft residual penalties can reduce such violations, but they do not generally ensure conservation or constraint satisfaction to numerical precision, particularly in meshed networks where loop-induced coupling makes feasibility highly topology-dependent~\cite{LuPestourie2021HardPINN,Mohan2023HardConstraints}. At the same time, many off-the-shelf message-passing surrogates are formulated mainly as node-state predictors, whereas steady-state gas hydraulics is intrinsically edge-mediated: pipe flows and pressure drops are defined on edges and coupled through nodal balance and network topology. These limitations leave a practical gap for surrogates that combine pipe-level representation, hard enforcement of nodal mass conservation, topologically consistent pressure reconstruction, and planning-oriented validation of feasibility margins and limiting components within a single differentiable graph-based workflow.

To address this gap, this work proposes a physics-informed graph neural network surrogate for steady-state gas-network simulation and feasibility analysis. The model predicts squared-pressure differences and flows on network edges, aligning the learned representation with the governing pressure--flow relations. A differentiable projection layer enforces nodal mass conservation as a hard constraint, while a Laplacian-based reconstruction maps predicted pressure differences to globally consistent nodal pressures. This hybrid learning--physics formulation couples an edge-centric GNN with deterministic constraint operators, enabling fast forward simulation while preserving the physical structure required for planning-oriented assessment.

The main contributions are: (i) an edge-centric GNN architecture aligned with gas-flow physics; (ii) exact enforcement of nodal mass conservation through a differentiable projection layer; (iii) deterministic reconstruction of topologically consistent pressure fields from learned edge quantities; and (iv) a planning-oriented validation on loadability assessment, feasibility screening, and sensitivity-based bottleneck identification for GasLib benchmarks.

The remainder of the paper is organized as follows. Section 2 introduces the PI-GNN architecture, the training objective, and the feasibility-screening formulation. Section 3 evaluates predictive accuracy, physical consistency, and runtime performance on GasLib benchmark networks, including an ablation study and a loadability analysis. Section 4 discusses the observed accuracy--consistency trade-offs and the practical scope of the surrogate. Section 5 summarizes the main findings, limitations, and implications for modern energy-system planning.

\section{Methods}

The proposed framework is a hybrid learning--physics surrogate for repeated steady-state gas-network simulation and feasibility evaluation. Each operating scenario is encoded as a graph input: nodes represent junctions, supply points, or demand points and carry operating and boundary-condition features, while edges represent physical pipes and carry static attributes such as length, diameter, and roughness. Reference pressure and flow labels are generated with MYNTS during training and evaluation; after training, inference requires only the encoded operating scenario and fixed network attributes.

\begin{figure}[t]
  \centering
  \includegraphics[width=0.98\linewidth]{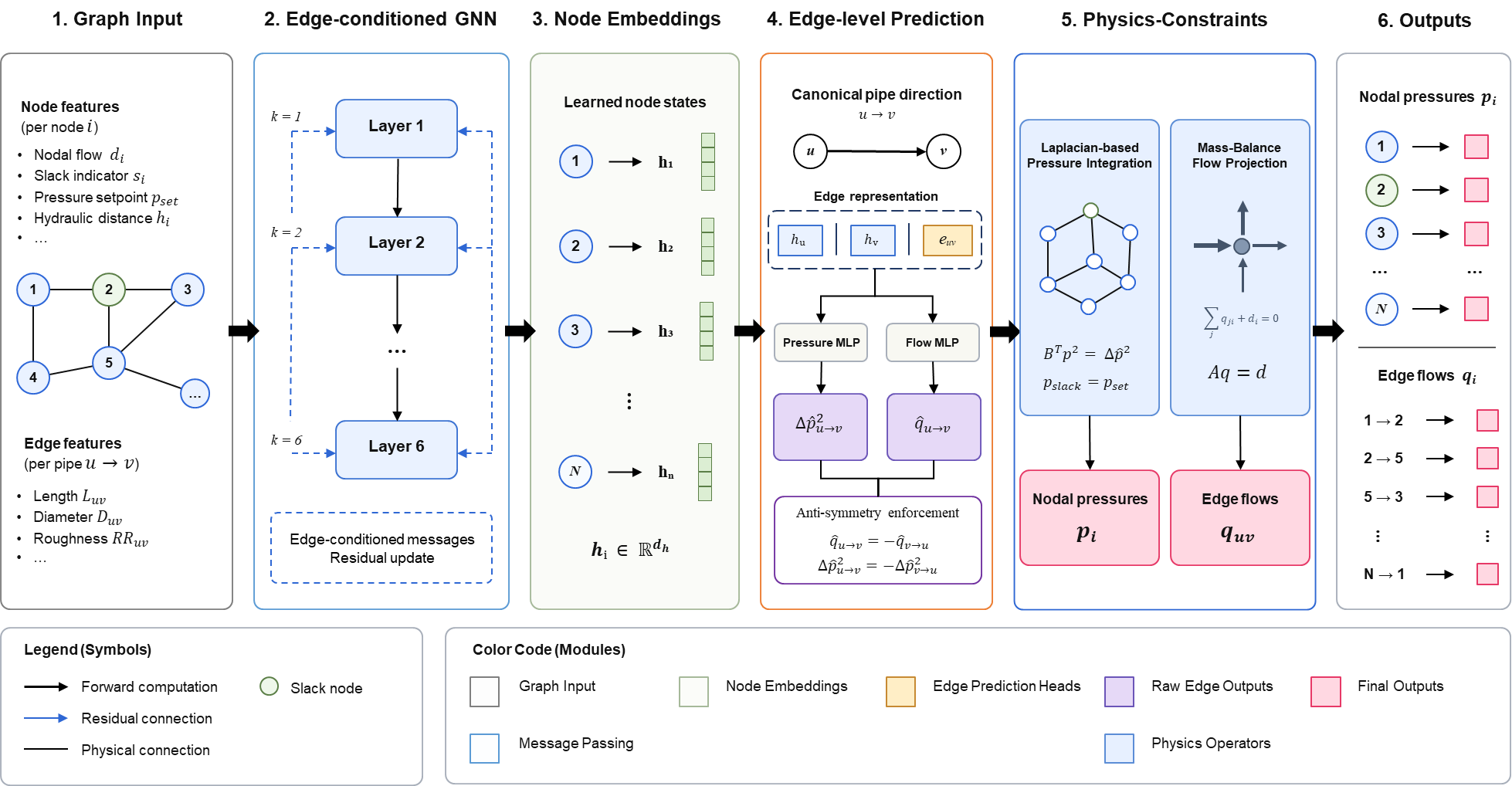}
  \caption{Architecture of the proposed physics-informed GNN surrogate for steady-state gas-network simulation. Each operating scenario is encoded as a directed graph with nodal operating and boundary-condition features and static pipe attributes. An edge-conditioned message-passing network predicts canonical-edge squared-pressure differences and preliminary pipe flows, with reverse-edge quantities imposed by antisymmetry. The physics-constrained reconstruction stage integrates pressure drops through a Laplacian solve with the reference pressure fixed and projects pipe flows onto the nodal mass-balance constraint, yielding topologically consistent pressures and mass-conserving flows for feasibility analysis.}
  \label{fig:surrogate_architecture}
\end{figure}

As summarized in Fig.~\ref{fig:surrogate_architecture}, the forward pass consists of six connected blocks: graph input, edge-conditioned message passing, learned node embeddings, edge-level prediction, physics-constrained reconstruction, and final outputs. The GNN maps the graph input to node embeddings, which are then combined with pipe attributes to predict canonical-edge squared-pressure differences and preliminary flows. Reverse-edge quantities are imposed by antisymmetry. Differentiable physics operators then reconstruct nodal pressures through a Laplacian solve and project edge flows onto the nodal mass-balance constraint.

The remainder of this section follows the same workflow. We first define the steady-state gas-flow problem and graph representation, then describe the edge-level GNN architecture, the pressure and flow reconstruction operators, the training loss, and finally the surrogate-based feasibility-screening and sensitivity-analysis procedure.

\subsection{Problem statement and hydraulic constraints}
\label{sec:method_overview}

We consider steady-state, isothermal gas flow on a network graph $G=(V,E)$, where nodes represent junctions, supply points, or demand points, and edges represent passive pipes. Each operating scenario $s$ specifies nodal injections or withdrawals $\mathbf{d}^{(s)}\in\mathbb{R}^N$ and one pressure-controlled reference node (slack node) $r^{(s)}$ with prescribed pressure setpoint $p_{\mathrm{set}}^{(s)}$. The sign convention is negative for supply and positive for demand. All generated scenarios satisfy global mass balance,

\begin{equation}
\sum_{i \in V} d_i^{(s)} = 0,\label{eq:global_mass}
\end{equation}
so the nodal operating conditions are balanced before they are passed to the surrogate.

For graph-based prediction, each physical pipe $\{u,v\}\in E$ is assigned a canonical orientation $e=(u\to v)$. Directed quantities on the reverse orientation are obtained by antisymmetry,
$q_{v\to u}=-q_{u\to v}$ and $\Delta p^2_{v\to u}=-\Delta p^2_{u\to v}$,
so a single physical pipe state is represented consistently in both directions. We use the pressure-drop convention $\Delta p^2_{u\to v}=p_u^2-p_v^2$, so $\Delta p^2$ has the same sign as $q$ for dissipative passive pipes. The pressure-controlled reference node fixes the pressure level through

\begin{equation}
p_{r^{(s)}}^{2(s)} = p_{\mathrm{set}}^{2(s)}.\label{eq:slack_bc}
\end{equation}
Squared-pressure variables are used because steady-state gas-pipeline relations are naturally expressed in pressure-squared differences~\cite{Osiadacz1987GasNetworks,Menon2005GasPipelineHydraulics}.

Let $\mathbf{A}\in\mathbb{R}^{N\times M}$ denote the node--edge incidence matrix of the physical pipe network under the canonical orientation, and let $\mathbf{q}\in\mathbb{R}^M$ be the corresponding vector of pipe flows. Nodal mass conservation is written as
\begin{equation}
\mathbf{A}\mathbf{q} = \mathbf{d},\label{eq:mass_physical}
\end{equation}
which links pipe flows to the prescribed nodal injections and withdrawals.

Along a passive pipe, steady-state momentum balance relates pressure loss to flow through a nonlinear friction law. In Weymouth-type form,
\begin{equation}
  p_u^{2(s)} - p_v^{2(s)} = R_{uv}\, q_{u\to v}^{(s)}\,\left|q_{u\to v}^{(s)}\right|,\label{eq:weymouth_generic}
\end{equation}
where $R_{uv}$ is a pipe-specific resistance coefficient determined by static attributes such as length, diameter, roughness, and the friction-factor correlation used by the reference simulator~\cite{Menon2005GasPipelineHydraulics,Thorley2004}.
Here $p_u^2-p_v^2$ denotes the squared-pressure drop for positive flow from $u$ to $v$, and we use this same convention for the directed edge variable $\Delta p^2_{u\to v}$.

\subsection{Scenario-to-graph encoding}
\label{sec:input_feature}

Each operating scenario is encoded on the fixed physical network topology. Scenario-dependent quantities are assigned to nodes, while static pipe properties are assigned to edges. This separates changing operating conditions from invariant network structure.

\paragraph{Node features.}
Each node $i\in V$ receives features describing its local operating condition and pressure-boundary role. The nodal injection or withdrawal $d_i^{(s)}$ is encoded as
\begin{equation}
x_i^{\mathrm{flow}} = \mathrm{asinh}\left(\frac{d_i^{(s)}}{q_0}\right),\label{eq:asinh}
\end{equation}
where $q_0$ is a characteristic flow scale computed from the training set as the median absolute nonzero edge-flow magnitude. The inverse hyperbolic sine transform preserves the sign of supply and demand while compressing large values; here $\mathrm{asinh}(x)=\ln\left(x + \sqrt{x^2 + 1}\right)$. Additional node features include indicators for the pressure-controlled reference node and pressure-setpoint location, a normalized topological distance to the reference node, and a log-normalized squared pressure setpoint broadcast to all nodes. In the reported configuration, four Laplacian positional encoding components are also appended.
These global reference features are included because pressure reconstruction is anchored at the pressure-controlled reference node, while message passing itself has a finite receptive field. They provide the GNN with coarse information about each node's position relative to the pressure boundary without requiring excessively deep message-passing stacks.

\paragraph{Edge features.}
Each physical pipe $\{u,v\}\in E$ is described by static attributes that affect hydraulic resistance: length $L_{uv}$, diameter $D_{uv}$, and roughness $\varepsilon_{uv}$. These quantities are converted to meters, log-transformed with a small positive offset, and standardized using training-set statistics. The same normalized pipe-feature vector is assigned to both directed representations of a physical pipe.

\paragraph{Output representation and scaling.}
Pressure-related quantities are represented in squared-pressure form. The GNN predicts normalized squared-pressure differences, which are rescaled by $c_{p^2}$, a high-percentile scale of training-set deviations $|p_i^2-p_{\mathrm{set}}^2|$. Flow is predicted through an asinh-normalized variable and mapped back to physical units by $q=q_0\sinh(\widetilde{q})$, so the final flow variables remain in Nm$^3$/s.

\subsection{Edge-centric GNN surrogate}
\label{sec:model_structure}

The surrogate predicts hydraulic quantities on edges, consistent with the fact that pipe flows and pressure losses are defined on pipes. A bidirected graph is used during message passing so that information can propagate in both directions along each physical pipe. Final predictions are made only once per pipe on a canonical orientation, and reverse-direction quantities are obtained by antisymmetry.

Node embeddings are computed with a stack of GINEConv message-passing layers~\cite{Gilmer2017NeuralMessagePassing,xu2019how,hu2020strategies}. The input node features are first mapped to hidden states \(\mathbf{h}_i^{(0)}\). The normalized pipe feature vector \(\mathbf{e}_{uv}\), containing transformed length, diameter, and roughness, is separately encoded as \(\boldsymbol{\eta}_{uv}\) so that it has the same hidden dimension as the node states. At layer \(k\), the update is

\begin{equation}
\mathbf{h}_i^{(k+1)}
=
\mathrm{MLP}^{(k)}\!\left[
(1+\epsilon^{(k)})\,\mathbf{h}_i^{(k)}
+
\sum_{j \in \mathcal{N}(i)}
\mathrm{ReLU}\!\left(\mathbf{h}_j^{(k)} + \boldsymbol{\eta}_{ji}\right)
\right],
\label{eq:mp_update}
\end{equation}
where \(\epsilon^{(k)}\) is learned. This update lets pipe attributes condition how neighboring node states are aggregated.

For each canonical pipe direction $(u\to v)$, the pressure-drop head receives the concatenation of the two endpoint embeddings and the encoded pipe feature. Using the convention $\Delta p^2_{u\to v}=p_u^2-p_v^2$, the model predicts
\begin{equation}
\begin{aligned}
  \widehat{\Delta p^2}_{u\to v} &= c_{p^2}\,
  g_{\theta}\!\left(\mathbf{h}_u,\mathbf{h}_v,\boldsymbol{\eta}_{uv}\right),\\
  \widehat{\Delta p^2}_{v\to u} &= -\widehat{\Delta p^2}_{u\to v},\\
  \widetilde{q}_{u\to v} &= f_{\theta}\!\left(\mathbf{h}_u,\mathbf{h}_v\right),\\
  \widehat{q}^{\mathrm{raw}}_{u\to v} &= q_0\sinh\!\left(\widetilde{q}_{u\to v}\right),\\
  \widehat{q}^{\mathrm{raw}}_{v\to u} &= -\widehat{q}^{\mathrm{raw}}_{u\to v}.
\end{aligned}
\label{eq:dp2_head}
\end{equation}
Here \(g_{\theta}\) and \(f_{\theta}\) are shared multilayer perceptrons, and \(c_{p^2}\) and \(q_0\) are the training-set scales defined in Section~\ref{sec:input_feature}. In the reported configuration, static pipe attributes enter the flow prediction through the message-passing embeddings, while the pressure-drop head uses the encoded pipe feature directly.

Because message passing has a finite receptive field, long-range hydraulic context is also supplied through the node features, including pressure-controlled reference-node indicators, the broadcast pressure setpoint, normalized topological distance, and Laplacian positional encodings.

\subsection{Physics-constrained pressure and flow reconstruction}
\label{sec:physics_layers}
The edge-level GNN outputs are local predictions and do not, by themselves, guarantee a globally valid hydraulic state. The reconstruction stage therefore applies two deterministic operators: pressure integration from predicted squared-pressure differences, and flow projection onto the nodal mass-balance constraint. The equations below are written for the canonical physical edge set; the implementation applies the same operations to the bidirected representation with reverse-edge antisymmetry.

Let $A\in\mathbb{R}^{N\times M}$ denote the incidence matrix under the canonical pipe orientation, with $-1$ at the tail node $u$ and $+1$ at the head node $v$. Then for $e=(u\to v)$,
$(A^\top\mathbf{p}^2)_e=p_v^2-p_u^2$, so the squared-pressure drop is $\Delta p^2_{u\to v}=p_u^2-p_v^2=-(A^\top\mathbf{p}^2)_e$. Nodal squared pressures are reconstructed by solving
\begin{equation}
\widehat{\mathbf{p}}^{2(s)}
=
\arg\min_{\mathbf{p}^{2}}\,\left\|-A^\top \mathbf{p}^{2} - \widehat{\boldsymbol{\Delta p^2}}^{(s)}\right\|_2^2
\quad \text{s.t. } p_{r^{(s)}}^2 = p_{\mathrm{set}}^{2(s)}.
\end{equation}
Equivalently, defining $\boldsymbol{\delta}^{(s)}=\mathbf{p}^{2(s)}-p_{\mathrm{set}}^{2(s)}\mathbf{1}$ with $\delta_{r^{(s)}}=0$, the free-node variables satisfy
\begin{equation}
A_{\mathcal{F}}A_{\mathcal{F}}^\top \boldsymbol{\delta}^{(s)}_{\mathcal{F}}
=
- A_{\mathcal{F}}\widehat{\boldsymbol{\Delta p^2}}^{(s)},
\qquad
\mathcal{F}=V\setminus\{r^{(s)}\},
\end{equation}
where \(A_{\mathcal{F}}\) contains the incidence rows of all non-reference nodes. This graph Laplacian solve projects the predicted edge differences onto an integrable pressure field with the reference pressure fixed~\cite{Chung1997,Spielman2007}.

Flow consistency is enforced by projecting the raw flow prediction \(\widehat{\mathbf{q}}^{\mathrm{raw}(s)}\) onto the affine space of mass-conserving flows:
\begin{equation}
\widehat{\mathbf{q}}^{(s)}
=
\arg\min_{\mathbf{q}}\,\left\| \mathbf{q} - \widehat{\mathbf{q}}^{\mathrm{raw}(s)} \right\|_2^2
\quad \text{s.t. } A\mathbf{q} = \mathbf{d}^{(s)}.
\end{equation}
The redundant balance equation at the pressure-controlled reference node is omitted from the solve. The KKT conditions give
\begin{equation}
A_{\mathcal{F}}A_{\mathcal{F}}^\top \boldsymbol{\lambda}^{(s)}_{\mathcal{F}}
=
A_{\mathcal{F}}\widehat{\mathbf{q}}^{\mathrm{raw}(s)}-\mathbf{d}^{(s)}_{\mathcal{F}},
\qquad
\widehat{\mathbf{q}}^{(s)}
=
\widehat{\mathbf{q}}^{\mathrm{raw}(s)}-A_{\mathcal{F}}^\top\boldsymbol{\lambda}^{(s)}_{\mathcal{F}}.
\end{equation}
Both operators are sparse linear solves and are differentiable with respect to the GNN outputs~\cite{AmosKolter2017OptNet,BoydVandenberghe2004}. They enforce pressure integrability and nodal mass conservation exactly, while the nonlinear pressure--flow relation is encouraged through the loss terms described next. This separation keeps the reconstruction step lightweight and deterministic, but it also means that pressure--flow compatibility must be evaluated and regularized rather than assumed.

\subsection{Training objective and reference-state supervision}
\label{sec:loss}

Training combines supervised agreement with the MYNTS reference solutions and soft physics-guided regularization of pressure--flow directionality~\cite{Karpatne2017,Willard2022ACMSurvey}. The hard operators in Section~\ref{sec:physics_layers} already enforce pressure integrability and nodal mass conservation. The loss function therefore focuses on matching the reference hydraulic state while discouraging residual inconsistencies between reconstructed pressures and projected flows.

The active objective is
\begin{equation}
\mathcal{L}
=
\lambda_{\Delta p^2}\mathcal{L}_{\Delta p^2}
+
\lambda_{p^2}\mathcal{L}_{p^2}
+
\lambda_{q}\mathcal{L}_{q}
+
\lambda_{\mathrm{mono}}\mathcal{L}_{\mathrm{mono}}
\, .
\end{equation}

The four terms supervise directed squared-pressure differences, reconstructed nodal squared pressures, pipe flows, and pressure--flow monotonicity, respectively. The corresponding weights and loss-configuration details are reported in Table~\ref{tab:appendix_training}. The implementation also supports additional turbulent-scaling and final-state pipe-consistency penalties, but their weights are set to zero in the reported experiments.

The pressure-drop loss is applied to the directed quantity $\Delta p^2_{u\to v}=p_u^2-p_v^2$, matching the sign convention used by pressure reconstruction. Predicted and reference values are normalized by $c_{p^2}$. To focus learning on hydraulically informative pipes, the main pressure-drop loss is evaluated on edges with large reference pressure differences, while a weak auxiliary penalty is retained on inactive physical edges. Nodal pressure supervision is applied after reconstruction, using squared-pressure deviations from the pressure-controlled reference value and excluding the fixed reference node.

Flow supervision is performed in the same transformed space used by the model output,
\begin{equation}
T_q(q)=\mathrm{asinh}\!\left(\frac{q}{q_0}\right).
\end{equation}
This improves conditioning while preserving sign. The flow loss combines supervision before and after mass-balance projection,
\begin{equation}
\mathcal{L}_{q}
=
\alpha \left\|T_q(\widehat{\mathbf{q}}^{\mathrm{raw}})-T_q(\mathbf{q})\right\|_1
+
(1-\alpha)\left\|T_q(\widehat{\mathbf{q}})-T_q(\mathbf{q})\right\|_1,
\end{equation}
where \(\alpha\in[0,1]\) balances learning of the raw GNN output and the final projected flow.

Finally, \(\mathcal{L}_{\mathrm{mono}}\) penalizes cases where the projected flow direction is inconsistent with the reconstructed pressure gradient. For a passive pipe, the desired relation is \(\widehat{q}_{u\to v}(\widehat{p}_u^2-\widehat{p}_v^2)\ge 0\). The penalty is applied only on edges with sufficiently large flow and pressure difference, so that near-zero-gradient cases do not dominate training.

\subsection{Planning-oriented feasibility screening and sensitivity analysis}
\label{sec:feasibility_screening}

\begin{figure}[t]
  \centering
  \includegraphics[width=0.75\linewidth]{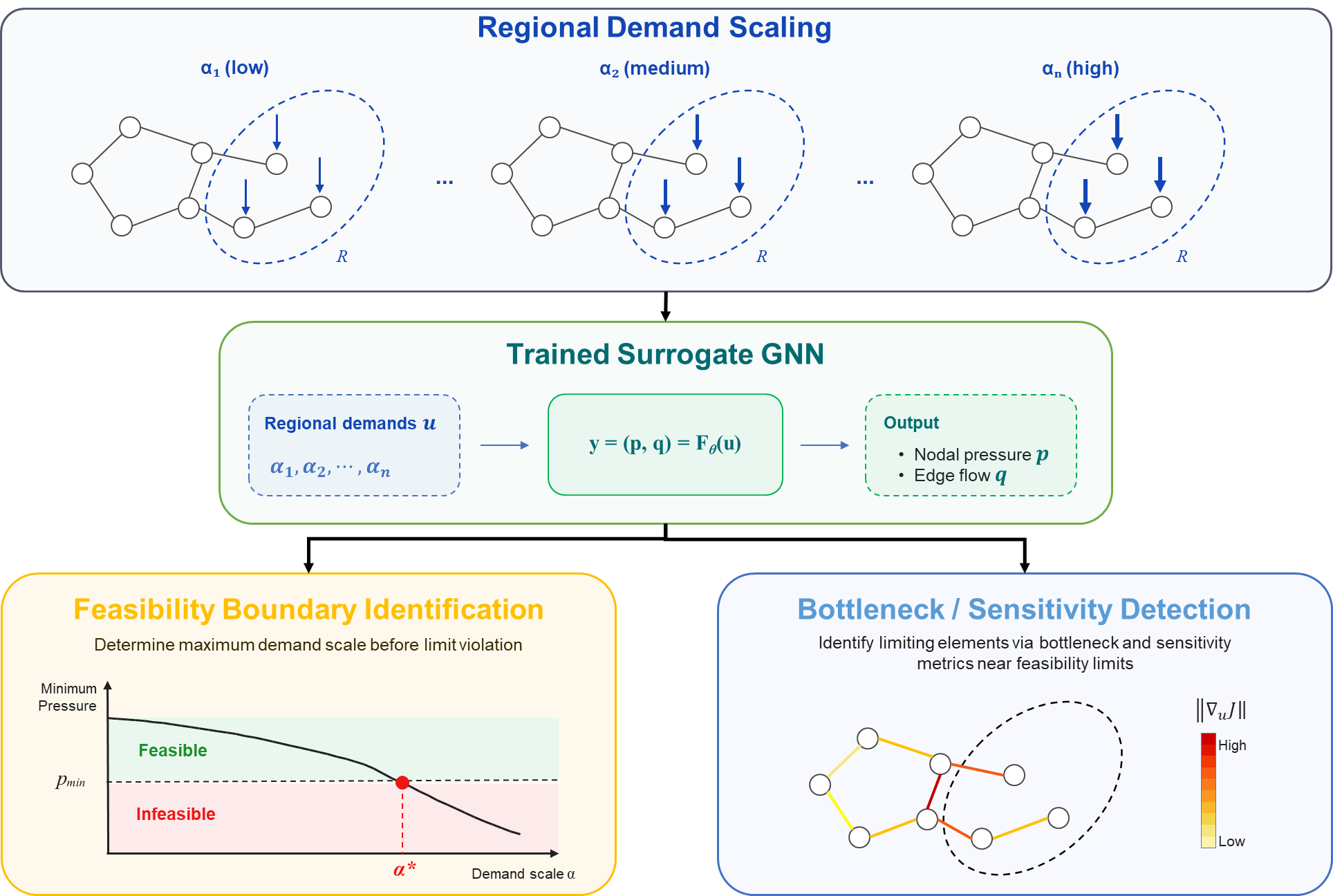}
  \caption{Surrogate-based demand-scaling and bottleneck-screening workflow. A regional demand set $\mathcal{R}$ is scaled by loading factors $\alpha$ while other boundary conditions are held fixed, and the trained surrogate $F_\theta$ predicts nodal pressures $\mathbf{p}$ and pipe flows $\mathbf{q}$. Feasibility is assessed from the minimum nodal pressure relative to the pressure limit $p_{\min}$, giving the critical loading factor $\alpha^*$. Near this boundary, bottleneck or sensitivity scores are computed for candidate pipes to identify components most associated with loss of feasibility.}
  \label{fig:feasibility_framework}
\end{figure}

After training, the surrogate provides a fast differentiable map from an operating scenario to reconstructed pressures and mass-conserving flows. This section uses that map for a planning-oriented task: estimating how far regional demand can be increased before pressure feasibility is lost, and identifying network components most associated with that limit. The task is intended as a controlled screening problem that can be validated against the reference simulator, rather than as a replacement for full operational optimization. The workflow is summarized in Fig.~\ref{fig:feasibility_framework}.

Let $\mathbf{d}_0$ denote a baseline balanced injection vector and let $\mathcal{R}$ be the target demand region. A scalar loading factor $\alpha$ is used to trace the feasibility boundary, analogous to loadability analysis in network planning~\cite{Ajjarapu2006}, and scales the withdrawals in this region,

\begin{equation}
d_i(\alpha) = \alpha d_{0,i}, \quad i \in \mathcal{R},
\end{equation}
while the remaining boundary conditions are held fixed and the balancing supply adjusts consistently with the scenario construction. The trained surrogate then predicts the reconstructed hydraulic state,
\begin{equation}
(\widehat{\mathbf{p}}(\alpha),\widehat{\mathbf{q}}(\alpha)) = F_\theta(\mathbf{u}(\alpha)),
\end{equation}
where $\mathbf{u}(\alpha)$ denotes the graph input features for the scaled scenario.

Feasibility is evaluated through the minimum nodal pressure. For a prescribed pressure limit $p_{\min}$, the critical loading factor is
\begin{equation}
\alpha^* = \max_{\alpha}\ \alpha
\quad \text{s.t.}\quad
\min_i \widehat{p}_i(\alpha) \geq p_{\min}.
\end{equation}
In the experiments, $\alpha^*$ is found by an expansion stage followed by binary search, using the monotonic decrease of minimum pressure observed along the tested demand-scaling paths.

The same differentiable surrogate is also used for bottleneck interpretation near the feasibility boundary. Scalar pressure-margin objectives can be differentiated with respect to input demand features or intermediate edge-level variables, producing sensitivity scores for candidate pipes or operating regions. These scores complement the forward feasibility scan by indicating which parts of the network have the strongest influence on the limiting pressure margin.

In addition to IID test scenarios, the evaluation includes stress-test scenario sets with increased total loading and regionally concentrated demand to assess whether the feasibility-screening workflow remains reliable under distribution shifts relevant to planning studies.

\section{Results}
\subsection{Experimental setup}
\label{sec:exp_setup}

Three GasLib benchmark transmission networks were used to evaluate the surrogate under controlled changes in topology and scale: GasLib-134, GasLib-135, and GasLib-582~\cite{Martin2015GasLib}. GasLib-134 is a purely branched network with no loops, whereas GasLib-135 and GasLib-582 contain loop structures that introduce additional flow-allocation degrees of freedom under nodal mass conservation. GasLib-582 provides the largest test case and is used for the detailed physical-consistency, ablation, and feasibility-screening analyses.

Table~\ref{tab:gaslib-benchmark} summarizes the structural characteristics of the benchmark networks and the sampled pressure-controlled reference setpoint range. The reported pressure range refers to the scenario-level reference pressure $p_{\mathrm{set}}^{(s)}$ at the pressure-controlled node, not to the minimum or maximum pressure observed across all nodes in the solved network state. Realized nodal pressures may therefore extend outside this setpoint interval depending on demand distribution, hydraulic losses, and network topology.

\begin{table}[width=0.95\linewidth,cols=5,pos=h]
\centering
\caption{Structural characteristics of the GasLib benchmark networks and sampled pressure-controlled reference setpoint ranges used for scenario generation.}
\label{tab:gaslib-benchmark}
\begin{tabular*}{\tblwidth}{@{\extracolsep{\fill}}lrrlc}
\hline
Network & Nodes & Pipes & Topology & Sampled $p_{\mathrm{set}}^{(s)}$ (bar)\\
\hline
GasLib-134 & 134 & 133 & Branched & 25--60\\
GasLib-135 & 135 & 170 & Meshed & 20--80\\
GasLib-582 & 582 & 609 & Meshed & 20--80\\
\hline
\end{tabular*}
\end{table}

Steady-state operating scenarios were generated on each fixed network topology. For each scenario $s$, the nodal injection vector $\mathbf{d}^{(s)}$, pressure-controlled reference node $r^{(s)}$, and reference pressure setpoint $p_{\mathrm{set}}^{(s)}$ were varied following the scenario-construction procedure described in Section~\ref{sec:input_feature}. Total demand was scaled and distributed across demand nodes using Gamma-based normalized weights, while the balancing supply was adjusted so that global mass balance was satisfied before simulation.

Each sampled scenario was solved with MYNTS~\cite{Clees2016MYNTS}, and only feasible steady-state solutions were retained. A pressure-span filter was applied to remove nearly uniform pressure profiles and retain scenarios with informative hydraulic variation.

For each network, training datasets with 1000, 2000, and 5000 feasible scenarios were generated. Each dataset was split into training and validation subsets using an 80/20 split. Model performance was evaluated on an independent test set of 500 feasible scenarios generated with the same procedure but not used during training or validation. All experiments were repeated over three independent random seeds. A compact summary of the scenario-generation and evaluation setup is provided in Table~\ref{tab:appendix_data}.

Unless otherwise noted, network characteristics are reported according to official GasLib specifications. Minor discrepancies in node or edge counts may arise due to preprocessing steps required for solver compatibility but do not affect the underlying physical structure or comparative analysis.

The evaluation considers predictive accuracy, physical consistency, and computational efficiency. Predictive accuracy is measured using pressure and flow MAE together with $R^2$. Physical consistency is assessed through nodal mass-balance residuals, pressure--flow monotonicity violations, and turbulent friction-factor behavior. Computational efficiency is measured as average inference time per scenario. Metric definitions are summarized in Table~\ref{tab:appendix_metrics}.

\subsection{Surrogate prediction accuracy}
\label{sec:surrogate_accuracy}

The predictive accuracy of the surrogate was evaluated using nodal pressure and pipe-flow errors on independent test sets. Table~\ref{tab:surrogate-perf} reports the mean and standard deviation over three random seeds for each network and training-set size. Flow metrics are computed after the mass-balance projection layer, consistent with the final reconstructed state used in downstream analysis.

\begin{figure}[width=1.0\linewidth,cols=1,pos=h]
\centering
\includegraphics[width=0.85\linewidth]{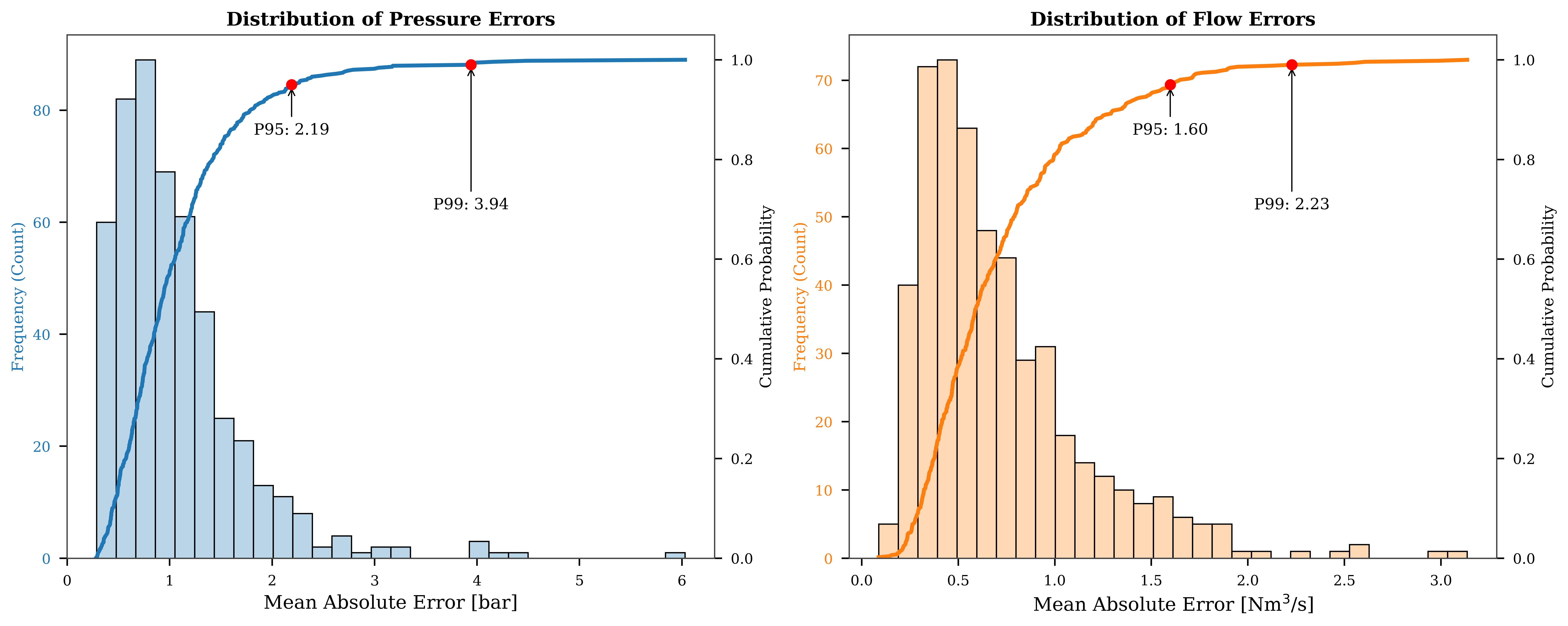}
\caption{Scenario-level pressure and flow error distributions for GasLib-582 using 5000 training scenarios and 500 independent test scenarios. Histograms show per-scenario MAE distributions, while overlaid curves show the corresponding CDFs. Red markers indicate the 95th and 99th percentiles.}
\label{fig:error-hist-cdf}
\end{figure}

\begin{figure}[width=1.0\linewidth,cols=1,pos=ht]
\centering
\includegraphics[width=0.9\linewidth]{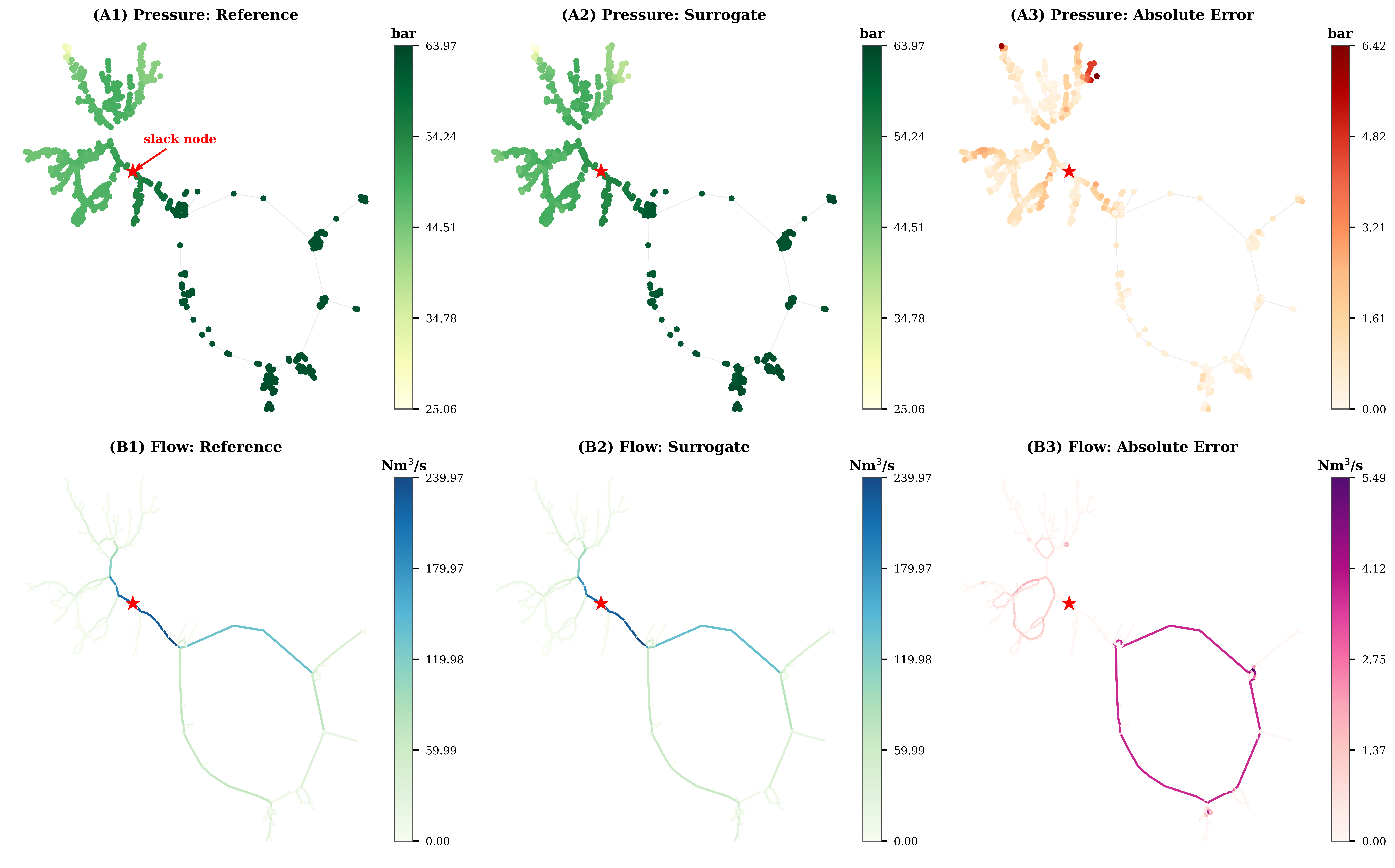}
\caption{Spatial distribution of pressure and flow predictions for a representative median-error GasLib-582 test scenario. Panels A1--A3 show reference pressure, surrogate pressure, and absolute pressure error at nodes (bar), while panels B1--B3 show reference flow, surrogate flow, and absolute flow error on pipes (Nm$^3$/s). The red star denotes the pressure reference node.}
\label{fig:spatial-errors-582}
\end{figure}

\begin{table}[width=0.95\linewidth,cols=6,pos=h]
\centering
\caption{Surrogate prediction performance across GasLib networks and training dataset sizes (mean $\pm$ std over 3 seeds).}
\label{tab:surrogate-perf}
\begin{tabular*}{\tblwidth}{@{\extracolsep{\fill}}lccccc}
\hline
Network & Training size & Pressure MAE (bar) & Pressure ($R^2$) & Flow MAE (Nm$^3$/s) & Flow ($R^2$)\\
\hline
\multirow{3}{*}{GasLib-134} & 1000 & 5.12 $\pm$ 1.32 & 0.627 $\pm$ 0.100 & $\sim 0$ & $\sim 1.000$\\
 & 2000 & 1.01 $\pm$ 0.20 & 0.898 $\pm$ 0.050 & $\sim 0$ & $\sim 1.000$\\
 & 5000 & 0.53 $\pm$ 0.10 & 0.939 $\pm$ 0.031 & $\sim 0$ & $\sim 1.000$\\
\multirow{3}{*}{GasLib-135} & 1000 & 2.59 $\pm$ 2.66 & 0.727 $\pm$ 0.378 & 0.80 $\pm$ 0.69 & 0.968 $\pm$ 0.028\\
 & 2000 & 0.81 $\pm$ 0.02 & 0.966 $\pm$ 0.002 & 0.99 $\pm$ 0.03 & 0.968 $\pm$ 0.002\\
 & 5000 & 0.21 $\pm$ 0.05 & 0.999 $\pm$ 0.001 & 0.82 $\pm$ 0.10 & 0.976 $\pm$ 0.005\\
\multirow{3}{*}{GasLib-582} & 1000 & 3.48 $\pm$ 1.45 & 0.891 $\pm$ 0.097 & 3.08 $\pm$ 0.14 & 0.967 $\pm$ 0.004\\
 & 2000 & 2.93 $\pm$ 0.29 & 0.934 $\pm$ 0.013 & 3.02 $\pm$ 0.19 & 0.966 $\pm$ 0.004\\
 & 5000 & 1.05 $\pm$ 0.02 & 0.981 $\pm$ 0.001 & 2.05 $\pm$ 0.01 & 0.972 $\pm$ 0.004\\
\hline
\end{tabular*}
\end{table}

Accuracy improves with increasing training-set size across the benchmark networks. For example, on GasLib-582, increasing the number of training scenarios from 1000 to 5000 reduces pressure MAE from 3.48 to 1.05~bar and increases pressure $R^2$ from 0.891 to 0.981. For scale, the final pressure MAE corresponds to approximately 1.3\% of the realized pressure range observed across the generated GasLib-582 dataset, which spans approximately 2--82~bar. This percentage is reported only as a dataset-level scale reference; the primary pressure-accuracy metrics are the absolute MAE and $R^2$ values.

The flow results should be interpreted together with the physics-constrained reconstruction. In GasLib-134, the network is purely branched, so for a prescribed $\mathbf{d}^{(s)}$ the mass-conserving pipe-flow solution is uniquely determined. The projection layer therefore maps raw flow predictions onto this unique feasible solution, giving near-zero flow error. This behavior reflects the interaction between topology and constraint enforcement rather than unconstrained model accuracy alone. In the meshed GasLib-135 and GasLib-582 networks, loop structures leave additional feasible flow degrees of freedom, so flow prediction depends on both learned edge-level information and projection. Even in these cases, projected flow predictions remain accurate, with flow $R^2$ values above 0.96 across all training sizes.

Pressure prediction is more sensitive to global reconstruction. The model predicts canonical-edge squared-pressure differences and reconstructs nodal pressures through the Laplacian solve described in Section~\ref{sec:physics_layers}. Local edge-level errors can therefore propagate along network paths before appearing as nodal-pressure error. This explains why pressure accuracy varies more strongly with topology and data coverage than projected flow accuracy. Figure~\ref{fig:error-hist-cdf} shows that most GasLib-582 test scenarios remain within a moderate pressure-error regime, with only a limited number of cases contributing to the upper error tail.

Figure~\ref{fig:spatial-errors-582} further indicates that larger pressure errors are spatially concentrated at nodes located farther from the pressure-controlled reference. This pattern is consistent with cumulative reconstruction effects along longer hydraulic paths in the network.

These results indicate that the surrogate reproduces pressure and flow behavior across both branched and meshed networks. The remaining pressure errors are primarily linked to global pressure reconstruction and limited training coverage, while projected flow predictions benefit from the embedded mass-balance constraint.

\subsection{Verification of physical consistency}
\label{sec:physical_consistency}

While the predictive accuracy results in Section~\ref{sec:surrogate_accuracy} demonstrate strong agreement between the surrogate and the reference solver, accuracy alone does not guarantee that predicted network states satisfy the governing physical constraints. These constraints arise from first principles of compressible pipeline flow—namely conservation of mass at network nodes and dissipative momentum balance along pipes, typically expressed through nonlinear pressure–flow relations (e.g., Weymouth-type equations). Ensuring consistency with these principles is essential not only for physical interpretability but also for the validity of downstream optimization and planning tasks.

The proposed framework enforces physical consistency at multiple levels with varying degrees of strictness: mass conservation is imposed as a hard constraint through projection, while pressure–flow monotonicity and turbulent scaling consistency are encouraged through physics-informed regularization and post-processing. In this section, we evaluate compliance with three key properties: (i) nodal mass conservation, (ii) pressure–flow monotonicity, and (iii) consistency with turbulent pipe-flow scaling laws.

\paragraph{Mass conservation in the learned flow representation.}

Nodal mass conservation constitutes a fundamental feasibility condition in steady-state gas network simulation, requiring that the net flow at each node matches the prescribed injection or withdrawal. Violations of this constraint lead to physically infeasible network states and render predictions unsuitable for planning or operational use.

In the proposed surrogate, mass conservation is incorporated directly into the model through a differentiable projection layer, as described in Section~\ref{sec:physics_layers}. Rather than treating conservation as an external correction, predicted edge flows are mapped onto the feasible subspace defined by the network incidence structure via a constrained projection derived from the KKT conditions. This operation is embedded within the model pipeline and participates in training, such that the surrogate learns flow representations that are consistent with mass balance after projection. As a result, the model operates in a constraint-aware output space where feasibility is ensured by construction.

\begin{figure}[width=1.0\linewidth,cols=1,pos=h]
\centering
\includegraphics[width=0.9\linewidth]{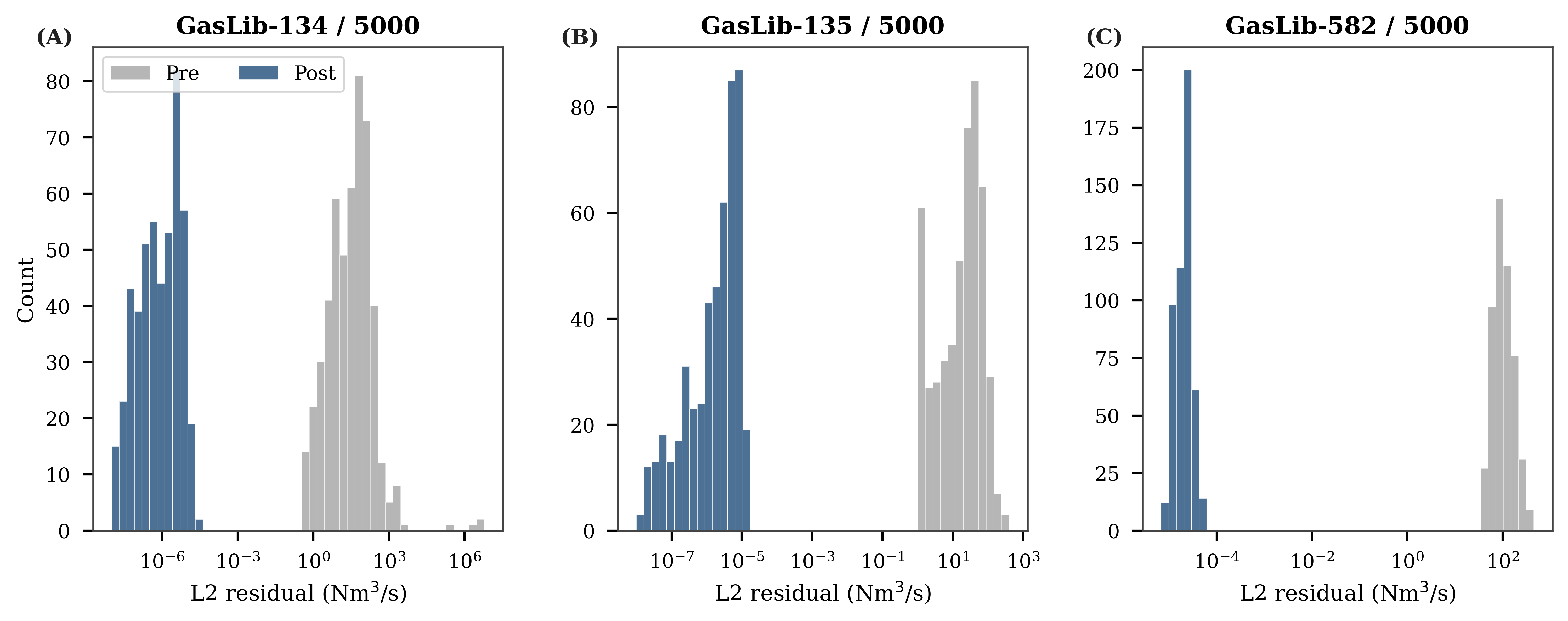}
\caption{Effect of the embedded mass-balance projection on nodal residuals for models trained with 5000 scenarios. For each GasLib network, gray bars show the pre-projection residuals and blue bars show post-projection residuals, where the per-scenario residual is the Euclidean norm $\|Aq-d\|_2$ in Nm$^3$/s. The logarithmic horizontal axis emphasizes the reduction from raw GNN imbalance to numerical-precision residuals after projection.}
\label{fig:mass-balance-dists}
\end{figure}

Figure~\ref{fig:mass-balance-dists} compares the distribution of nodal mass-balance residuals before (Pre) and after (Post) the projection layer. The residual is measured as the $\ell_2$ norm of nodal imbalance, $\|Aq - d\|_2$, computed per scenario. While raw predictions exhibit residuals on the order of $\mathcal{O}(10^{1}\text{–}10^{2})$ Nm$^3$/s, the projected outputs reduce these violations to $\mathcal{O}(10^{-5}\text{–}10^{-4})$ Nm$^3$/s, corresponding to solver-level numerical precision. This reduction is consistent across the full distribution, indicating that conservation is satisfied globally rather than selectively corrected for large violations. The isolated contribution of this constraint-aware layer is further examined in the ablation study in Section~\ref{sec:ablation}.

\paragraph{Monotonicity of pressure--flow relationships.}

For passive pipeline elements, steady-state gas flow is driven by pressure gradients and must proceed from higher to lower pressure, implying the monotonicity condition $q_{u\to v}(p_u^2 - p_v^2) \ge 0$. In the proposed surrogate, this condition is not enforced as a hard constraint but is encouraged through physics-informed loss terms and optional post-processing.

\begin{table}[width=0.95\linewidth,cols=5,pos=h]
\centering
\caption{Monotonicity violations under pressure-gradient thresholds for selected GasLib-582}
\label{tab:mono_true_threshold}
\begin{tabular*}{\tblwidth}{@{\extracolsep{\fill}}lrrrr}
\hline
Threshold $|\Delta p|$ (bar) & Edge Count & Retained (\%) & Violations & Violation Rate \\
\hline
All edges & 304500 & 100.0\% & 19655 & 6.45\% \\
$0<|\Delta p|<0.1$ & 239698 & 78.7\% & 14751 & 6.15\% \\
$0.1<|\Delta p|<0.5$ & 40718 & 13.4\% & 3862 & 9.48\% \\
$0.5<|\Delta p|<1.0$ & 9868 & 3.2\% & 139 & 1.41\% \\
$|\Delta p|>1.0$ & 4253 & 1.4\% & 1 & 0.02\% \\
\hline
\end{tabular*}
\end{table}

\begin{figure}[width=1.0\linewidth,cols=1,pos=h]
\centering
\includegraphics[width=\linewidth]{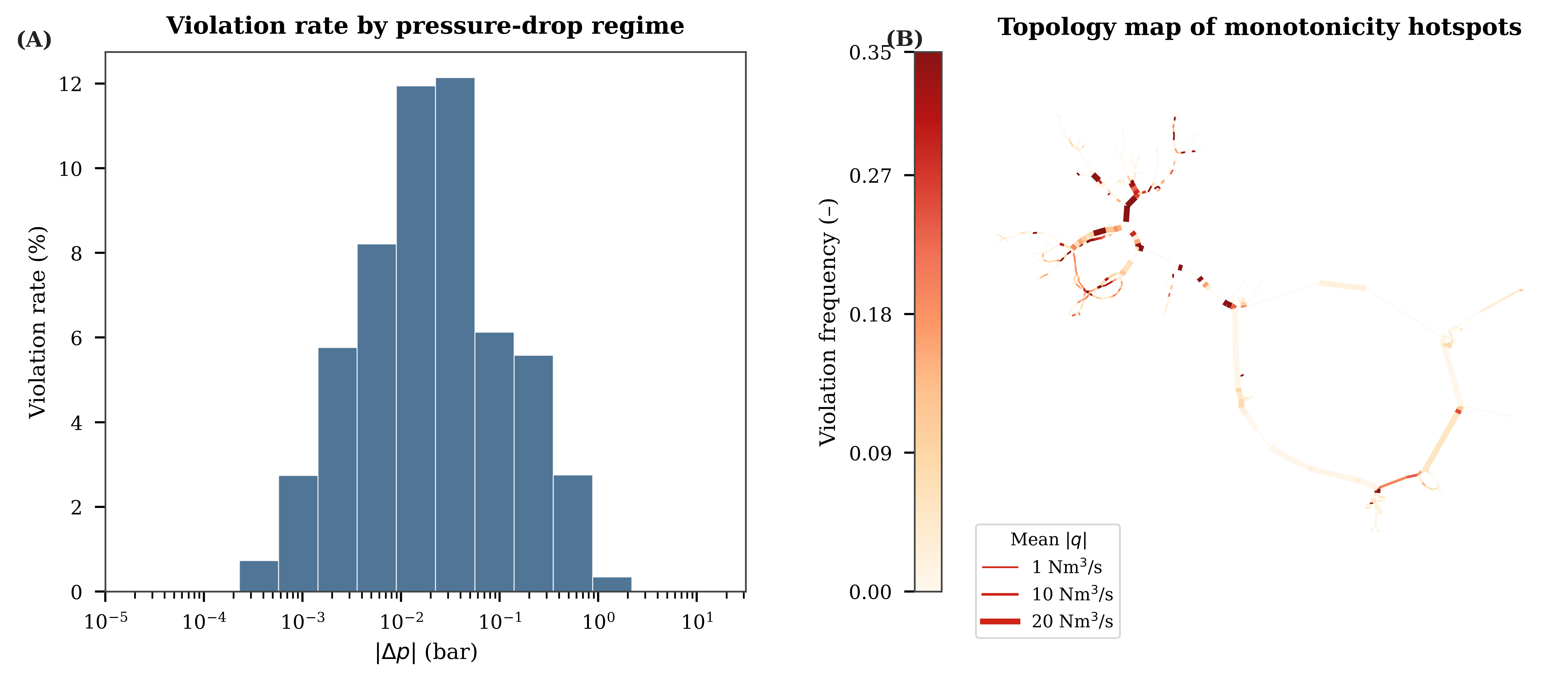}
\caption{Pressure--flow monotonicity diagnostics for GasLib-582 with 5000 training scenarios. (A) Violation rate as a function of pressure-drop magnitude $|\Delta p|$ (bar), where a violation denotes a pipe for which the predicted flow direction is inconsistent with the predicted pressure drop. (B) Network map of violation frequency, with color indicating the fraction of scenarios violating monotonicity on each pipe and line width indicating mean absolute flow $|q|$ (Nm$^3$/s). The figure distinguishes low-pressure-drop regimes from spatially localized recurrent violations.}
\label{fig:monotonicity-dists}
\end{figure}

The overall monotonicity violation rate is 6.45\%, computed across all edges and all validation scenarios for the GasLib-582 network. While this aggregate measure provides a global summary of directional inconsistency, Table~\ref{tab:mono_true_threshold} and Figure~\ref{fig:monotonicity-dists} show that violations are concentrated in regimes with relatively small pressure differences. Specifically, the violation rate decreases from 6.15\% for $0<|\Delta p|<0.1$ bar to 1.41\% for $0.5<|\Delta p|<1.0$ bar and only 0.02\% for $|\Delta p|>1.0$ bar, indicating that monotonicity is almost always preserved under sufficiently large pressure gradients. In low-pressure-gradient regimes, where the driving force for flow is weak, even minor prediction errors in either pressure or flow can lead to sign inconsistencies without materially affecting the overall system state.

While this pressure-gradient perspective indicates that violations are predominantly associated with relatively small pressure differences, a complementary analysis based on flow magnitude shows that violations are not confined to low-flow conditions. In particular, edges in the top 20\% of flow magnitude account for a substantial fraction of violations, indicating that even in high-flow regions, small pressure gradients combined with local prediction errors can lead to sign inconsistencies. This behavior reflects a structured limitation of the surrogate, in which strict monotonicity is not enforced but emerges indirectly from the learning objective, rather than a consequence of random fluctuations or numerical instability.



Despite these local inconsistencies, their impact on system-level behavior remains limited. As shown above, monotonicity violations are primarily associated with regimes of small pressure differences, where the driving force for flow is weak and directional ambiguity is inherently high. In such cases, sign inconsistencies typically correspond to small perturbations in pressure or flow and do not materially alter the overall hydraulic state. 

At the network level, the dominant pressure structure and flow distribution remain well captured, as evidenced by the strong predictive accuracy reported in Section~\ref{sec:surrogate_accuracy}. This indicates that the surrogate preserves the key physical relationships governing system behavior, even though strict monotonicity is not enforced. The observed violations therefore reflect a trade-off between model flexibility and strict physical consistency, arising from the use of soft constraints for differentiability.

While further reduction of monotonicity violations remains a potential direction for improvement, the current level of consistency is sufficient for planning-oriented applications, where global feasibility and system-level trends are of primary importance.

\paragraph{Consistency with turbulent friction-factor behavior.}

A further validation of physical consistency is obtained by examining whether the surrogate preserves the characteristic turbulent pipe-flow regime. Under steady-state conditions, gas transport in pipelines is governed by nonlinear friction laws that couple Reynolds number, pipe roughness, friction factor, and pressure loss~\cite{White2016}. In the squared-pressure formulation adopted in this work, this behavior is closely related to the expected nonlinear pressure--flow scaling and provides a useful diagnostic of admissible hydraulic states.

To assess this behavior, a classical Moody-type representation is employed. The diagram compares the inferred Darcy friction factor against Reynolds number for both MYNTS reference solutions and surrogate predictions, with the Colebrook--White relation shown as a reference trend. This diagnostic does not directly plot flow magnitude against squared-pressure drop; instead, it tests whether the surrogate-predicted states remain in the same turbulent friction regime as the reference simulator.

\begin{figure}[width=1.0\linewidth,cols=1,pos=h]
\centering
\includegraphics[width=\linewidth]{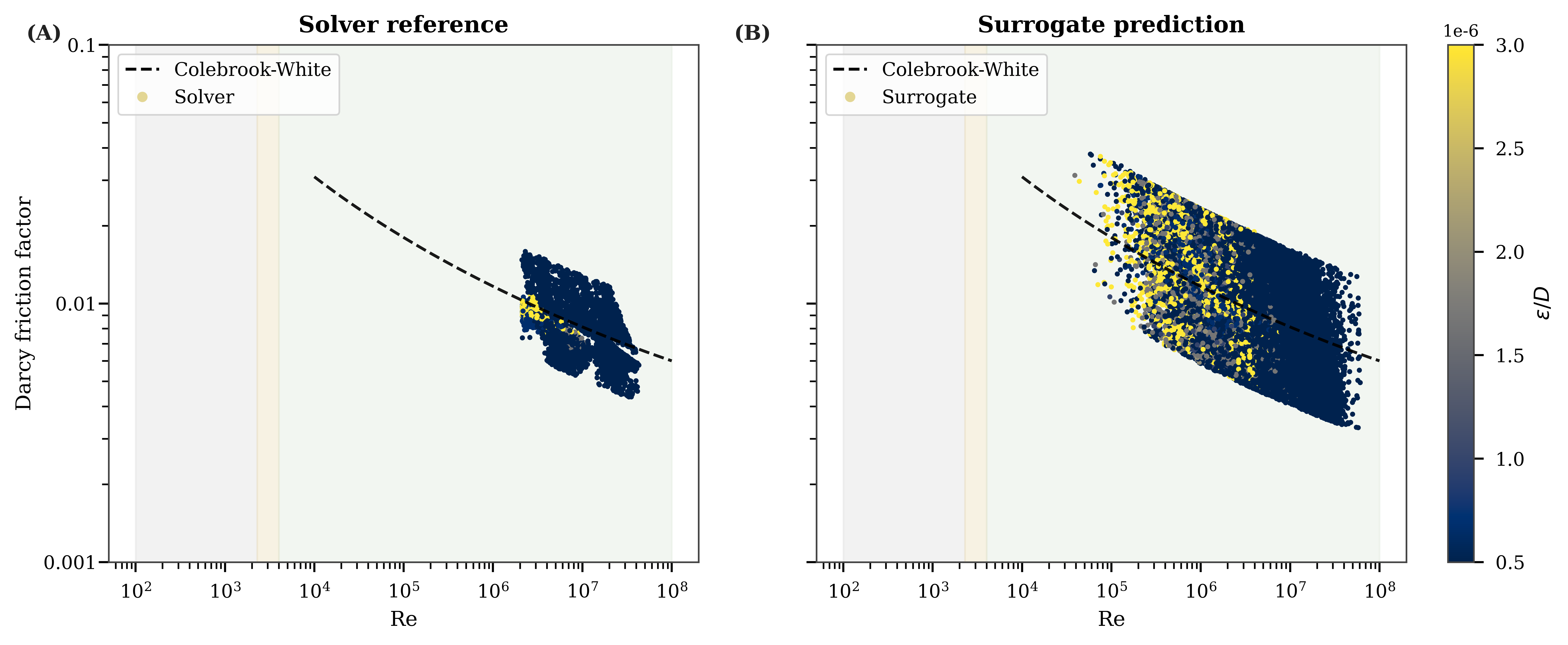}
\caption{Moody-type hydraulic consistency check for GasLib-582. Each point represents an inferred pipe-state pair from the MYNTS reference solutions (A) or surrogate predictions (B), plotted as Darcy friction factor versus Reynolds number $Re$; the dashed line is the Colebrook--White reference trend, point color encodes relative roughness, and the background shading marks Reynolds-number regime bands. Agreement in the occupied turbulent regime supports preservation of the dominant friction-factor behavior, while dispersion around the trend indicates surrogate approximation error.}
\label{fig:moody}
\end{figure}

Figure~\ref{fig:moody} compares surrogate predictions with reference solver (MYNTS) results under this representation. Both surrogate and solver outputs occupy the expected turbulent range and follow the same friction-factor trend, indicating that the dominant pipe-flow regime is preserved. Compared to the solver, the surrogate exhibits increased dispersion around the reference curve, reflected in a higher median deviation. This spread reflects approximation error inherent to the learned model rather than a systematic regime shift. Importantly, no systematic bias away from the expected Moody-type behavior is observed, suggesting that the surrogate maintains consistency with the governing pressure--flow physics despite local prediction errors.

\paragraph{Summary of physical consistency.}
The results establish a clear hierarchy of physical consistency within the proposed surrogate model. Mass conservation is strictly enforced through projection, achieving solver-level accuracy across all scenarios. Pressure--flow monotonicity is not guaranteed but is largely preserved, with violations concentrated in locally ambiguous regimes. The Moody-type analysis indicates that the surrogate remains consistent with the expected turbulent friction-factor regime despite increased dispersion.

These findings indicate that the surrogate achieves a balance between strict enforcement of essential feasibility constraints and flexible approximation of more complex physical relationships. While not fully physically exact, the model preserves the dominant structures governing system behavior, supporting its use in applications where feasibility and global consistency are primary requirements.

\subsection{Computational efficiency}
\label{sec:runtime}
In addition to predictive accuracy and physical consistency, a key motivation for surrogate modeling is the reduction in computational cost relative to conventional steady-state simulation. Classical gas network solvers evaluate nonlinear flow equations using iterative numerical methods (e.g., Newton-type schemes), which require repeated updates until convergence. While these approaches provide high-fidelity solutions, their computational cost becomes significant when large numbers of operating scenarios must be evaluated. The proposed surrogate replaces this iterative process with a direct forward evaluation of a physics-constrained model, in which the neural network and embedded constraint layers jointly produce the final solution.

Table~\ref{tab:runtime} reports the average per-scenario runtime for the surrogate and MYNTS. The surrogate produces the reconstructed pressure field and mass-conserving flow state in milliseconds, whereas MYNTS requires seconds per feasible steady-state solution under the same benchmark settings.

\begin{table}[width=0.95\linewidth,cols=5,pos=h]
\centering
\caption{Average surrogate inference runtime per scenario and comparison with MYNTS.}
\label{tab:runtime}
\begin{tabular*}{\tblwidth}{@{\extracolsep{\fill}}lrrrr}
\hline
Network & Nodes & Pipes & Surrogate (ms) & MYNTS (s)\\
\hline
GasLib-134 & 134 & 133 & 3.17 & $\sim 1.0$\\
GasLib-135 & 135 & 170 & 3.10 & $\sim 1.0$\\
GasLib-582 & 582 & 609 & 35.36 & $\sim 2.0$\\
\hline
\end{tabular*}
\end{table}

The surrogate runtime increases with network size. GasLib-134 and GasLib-135 require approximately 3~ms per scenario, while GasLib-582 requires 35.36~ms. This increase reflects the larger graph and the additional cost of the physics-constrained reconstruction operations, including Laplacian-based pressure reconstruction and mass-balance projection. Even so, the largest benchmark remains below 40~ms per scenario, while retaining the physical-consistency properties evaluated in Section~\ref{sec:physical_consistency}.

The benefit of millisecond-level inference is most relevant when steady-state simulation is used repeatedly rather than for a single operating point. A single MYNTS run taking one or two seconds may be acceptable for conventional planning checks. However, scenario-based studies can quickly require many thousands of evaluations. For example, hourly demand profiles over one year already give 8760 operating states; combining these with multiple demand-growth assumptions, supply allocations, pressure-setpoint choices, contingency cases, or regional stress tests can expand the workload to $10^5$--$10^6$ steady-state evaluations. At seconds per case, this becomes a batch-computing task. At millisecond-level inference, the same screening can be repeated interactively during sensitivity analysis, uncertainty exploration, or candidate-scenario filtering.

Thus, the surrogate is not intended to remove the need for validated solver-based verification. Its practical value is to shift the computational burden: large candidate sets can be screened rapidly with the surrogate, and MYNTS can then be reserved for final assessment of selected critical cases. This division is particularly useful for feasibility screening and bottleneck identification, where the objective is often to rank or prioritize operating conditions before detailed hydraulic confirmation.

\subsection{Baseline comparison and ablation analysis}
\label{sec:ablation}

\begin{figure}[width=\linewidth,cols=1,pos=h]
\centering
\includegraphics[width=\linewidth]{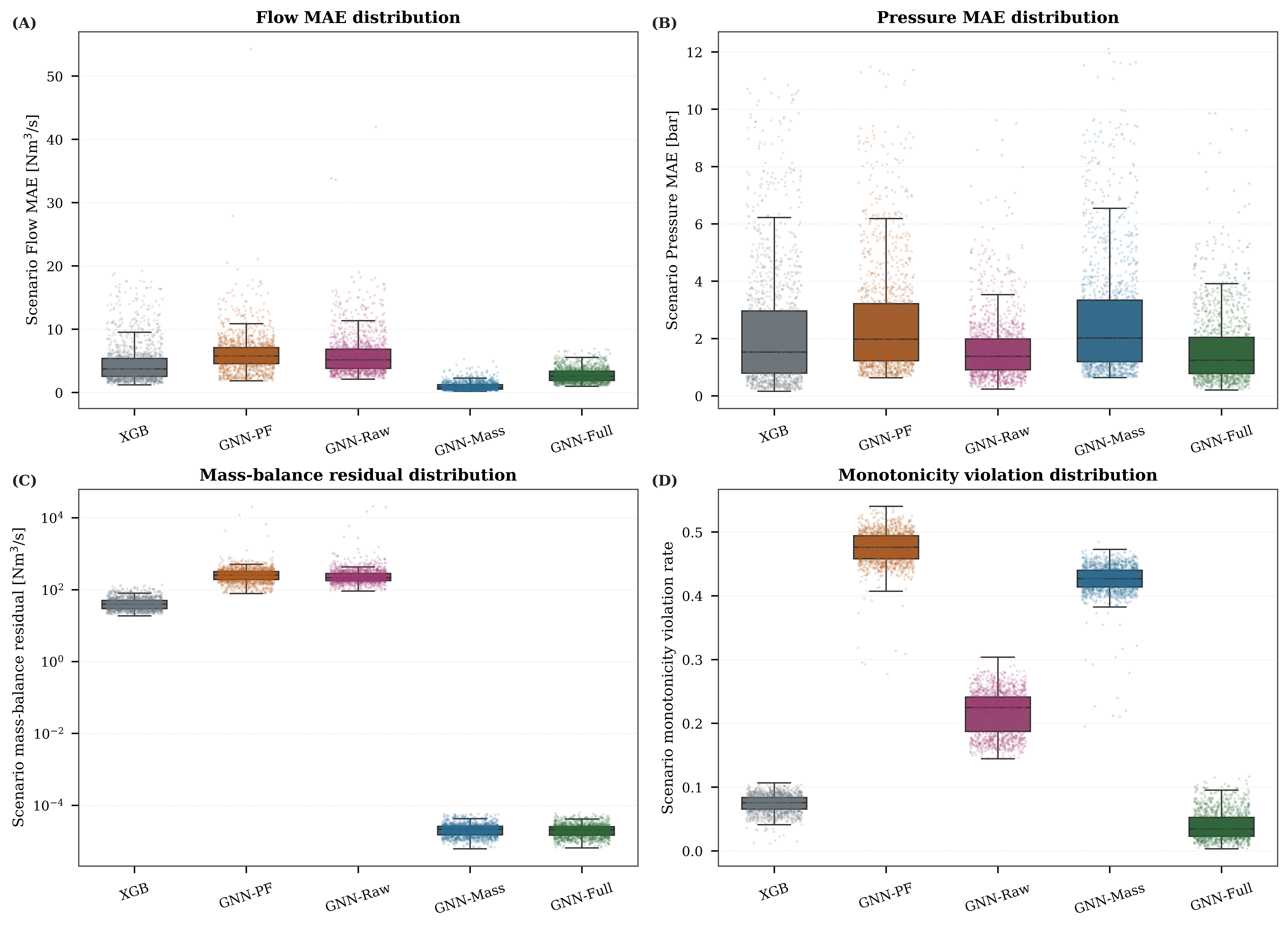}
\caption{Distributional ablation comparison on GasLib-582 test scenarios. Boxplots with overlaid scenario points compare XGB, GNN-PF, GNN-Raw, GNN-Mass, and GNN-Full for (A) flow MAE (Nm$^3$/s), (B) pressure MAE (bar), (C) mass-balance residual $\|Aq-d\|_2$ (Nm$^3$/s, logarithmic scale), and (D) monotonicity violation rate. The panels show whether differences in Table~\ref{tab:ablation} persist across the full scenario distribution rather than being driven by mean values alone.}
\label{fig:ablation-distribution}
\end{figure}

A controlled baseline and ablation study was conducted on the GasLib-582 network to isolate the respective roles of data-driven learning, graph-based representation, and physics-constrained reconstruction. The comparison includes a tabular XGBoost (XGB) baseline and four GNN variants with progressively stronger alignment to the governing physical structure: a pressure-first formulation (GNN-PF), an edge-first model without constraint layers (GNN-Raw), an edge-first model with embedded mass-conservation projection (GNN-Mass), and the full physics-constrained model (GNN-Full). Performance is evaluated using both predictive accuracy and physical-consistency metrics, as summarized in Table~\ref{tab:ablation}. The corresponding scenario-level distributions are shown in Fig.~\ref{fig:ablation-distribution}.

\begin{table}[width=0.98\linewidth,cols=7,pos=h]
\centering
\caption{Baseline comparison and physics-constrained ablation on GasLib-582 (mean $\pm$ standard deviation over three seeds).}
\label{tab:ablation}
\begin{tabular*}{\tblwidth}{@{\extracolsep{\fill}}lrrrrrr}
\hline
Model & \shortstack{Pressure\\MAE (bar)} & \shortstack{Pressure\\$R^2$} & \shortstack{Flow\\MAE (Nm$^3$/s)} & \shortstack{Flow\\$R^2$} & \shortstack{Mass Residual\\(Nm$^3$/s)} & \shortstack{Mono.\\Viol.}\\
\hline
XGB & $2.262 \pm 0.009$ & $0.959 \pm 0.001$ & $4.564 \pm 0.013$ & $0.899 \pm 0.001$ & $4.32\times 10^{1} \pm 0.63$ & $0.074 \pm 0.002$\\
GNN-PF & $2.563 \pm 0.052$ & $0.950 \pm 0.002$ & $6.105 \pm 0.277$ & $0.515 \pm 0.330$ & $2.95\times 10^{2} \pm 2.99\times 10^{1}$ & $0.476 \pm 0.022$\\
GNN-Raw & $1.634 \pm 0.088$ & $0.981 \pm 0.001$ & $5.829 \pm 0.670$ & $0.431 \pm 0.406$ & $2.91\times 10^{2} \pm 1.15\times 10^{2}$ & $0.218 \pm 0.034$\\
GNN-Mass & $2.642 \pm 0.051$ & $0.947 \pm 0.002$ & $0.965 \pm 0.089$ & $0.996 \pm 0.001$ & $2.22\times 10^{-5} \pm 2.51\times 10^{-7}$ & $0.426 \pm 0.014$\\
GNN-Full & $1.05 \pm 0.02$ & $0.981 \pm 0.001$ & $2.05 \pm 0.01$ & $0.972 \pm 0.004$ & $2.15\times 10^{-5} \pm 2.17\times 10^{-7}$ & $0.040 \pm 0.018$\\
\hline
\end{tabular*}
\end{table}

The XGB baseline provides a useful reference for fixed-topology regression. It achieves moderate pressure and flow accuracy, indicating that part of the steady-state mapping can be learned from tabular operating features. However, its outputs are not physically feasible: the mass-balance residual remains on the order of $10^{1}$ Nm$^3$/s, and monotonicity violations are present. This limits its use for planning workflows that require hydraulically admissible states, even when pointwise error metrics appear acceptable.

Introducing graph message passing without aligning the output formulation to pipe-level physics does not resolve this issue. GNN-PF gives pressure accuracy similar to XGB but performs poorly on flow prediction and physical-consistency metrics. This indicates that graph structure alone is not enough when the prediction targets are not matched to the edge-mediated nature of gas transport.

The edge-first raw model clarifies this point. GNN-Raw improves pressure prediction relative to XGB and GNN-PF, reaching a pressure MAE of 1.634~bar and pressure $R^2$ of 0.981. This supports the use of canonical-edge squared-pressure differences as the primary learned representation. However, without projection, the predicted flows remain infeasible, with mass residuals on the order of $10^{2}$ Nm$^3$/s and low flow $R^2$. Learning local edge-level hydraulic relationships therefore does not guarantee global mass conservation.

Adding the mass-balance projection changes the behavior substantially. GNN-Mass reduces mass residuals to numerical precision and gives the best projected flow metrics in Table~\ref{tab:ablation}. This outcome is expected because the projection directly constrains the feasible flow space, and the model variant does not include the full pressure-reconstruction and monotonicity structure. The improvement in flow accuracy should therefore be interpreted as a result of constraint enforcement, not as evidence that the pressure and flow fields are fully compatible. Indeed, GNN-Mass shows weaker pressure accuracy and a high monotonicity violation rate, indicating that conservation alone is insufficient to recover a coherent hydraulic state.

GNN-Full provides the most balanced result across the evaluated criteria. It preserves numerical-precision mass conservation, achieves the lowest pressure MAE among the compared models, and reduces monotonicity violations from 0.426 in GNN-Mass to 0.040. Its flow MAE is higher than GNN-Mass, but remains accurate in absolute terms and is accompanied by stronger pressure consistency. This trade-off is consistent with the role of the full reconstruction pipeline: the model is not optimized only for projected flow agreement, but for a final hydraulic state that balances pressure accuracy, mass conservation, and pressure--flow directionality.

\begin{figure}[width=\linewidth,cols=1,pos=h]
\centering
\includegraphics[width=0.8\linewidth]{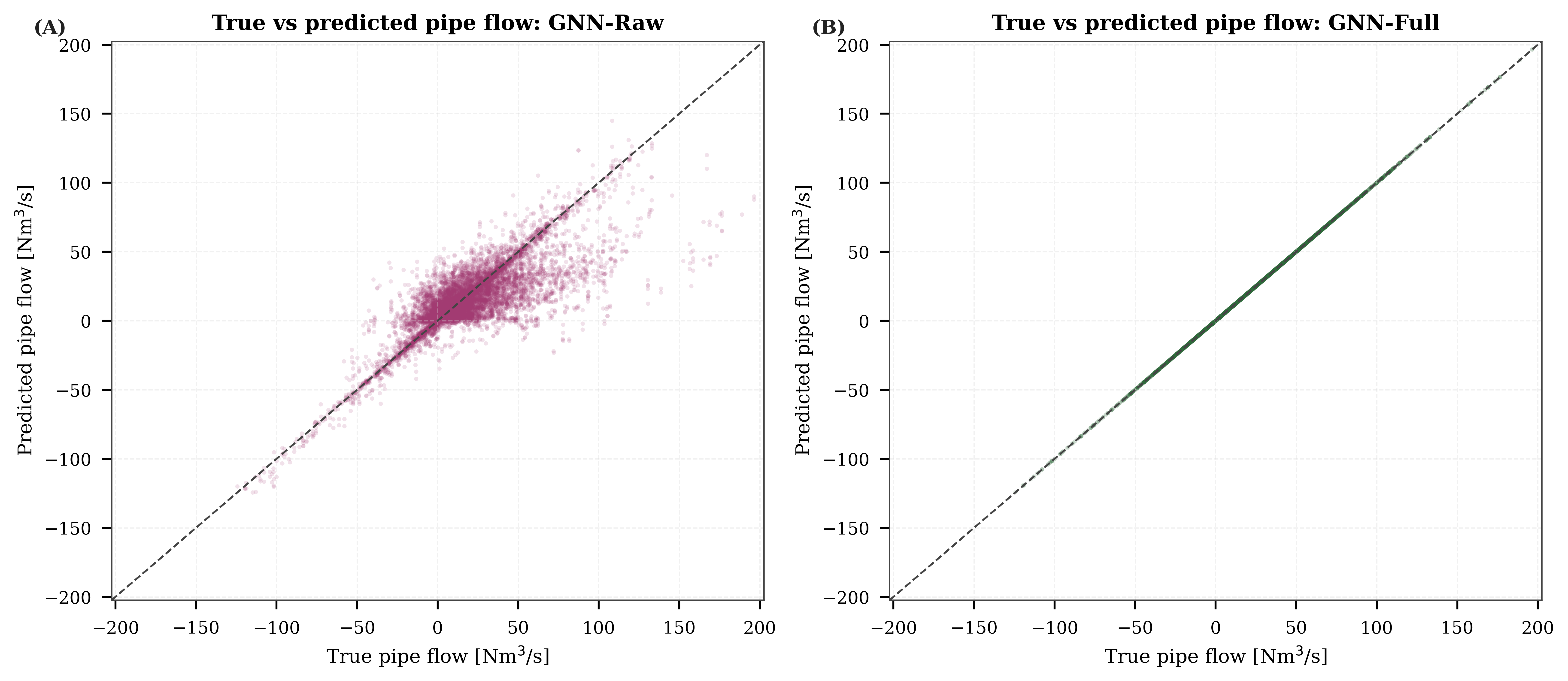}
\caption{Effect of mass-conservation projection on GasLib-134 pipe-flow predictions. Points compare true and predicted canonical pipe flows (Nm$^3$/s) against the one-to-one dashed line for (A) the unconstrained edge-first GNN output (GNN-Raw) and (B) the final physics-constrained model output (GNN-Full, after mass-balance projection and pressure reconstruction). Because GasLib-134 is nearly tree-like, enforcing nodal balance strongly constrains the feasible pipe-flow solution.}
\label{fig:projection-effect}
\end{figure}

The projection effect is especially clear on GasLib-134, shown in Figure~\ref{fig:projection-effect}. Because this network is purely branched, the mass-conserving pipe-flow solution is uniquely determined by the nodal injections. The projection therefore maps scattered raw edge predictions onto the feasible flow solution. This explains the near-zero flow errors observed for GasLib-134 in Table~\ref{tab:surrogate-perf} and reinforces the interpretation that topology and constraint enforcement jointly shape the final flow metrics.

Overall, the ablation results show that each component addresses a different failure mode. Edge-level prediction improves hydraulic representation, mass projection restores feasibility, and pressure reconstruction with monotonicity regularization improves coherence between pressure and flow. The full model is therefore preferred not because it is best on every individual metric, but because it provides the most reliable physically constrained state for downstream screening.

\subsection{Feasibility screening and bottleneck analysis}
\label{sec:results_screening}

To evaluate the surrogate in a planning-oriented inverse-use setting, a loadability screening study was conducted on the GasLib-582 network. Rather than solving a general high-dimensional optimization problem, the analysis focuses on a controlled feasibility screening task along a one-dimensional demand-scaling trajectory. This formulation enables direct validation against the reference solver while isolating the surrogate's ability to capture feasibility boundaries.

Five high-stress baseline scenarios were selected from the test set, including the maximum-demand case and four regionally dominated demand patterns (NW, NE, SW, SE). For each scenario, a scalar loading factor $\alpha$ was applied to a subset of demand nodes, while all other boundary conditions were held fixed. Feasibility was defined through a minimum-pressure constraint of 16~bar, and the critical loading factor $\alpha^*$ was determined using an expansion stage followed by binary search.

\begin{table}[width=0.98\linewidth,cols=8,pos=h]
\centering
\caption{Surrogate-based loadability screening results on GasLib-582 across representative high-demand scenarios.}
\label{tab:loadability}
\begin{tabular*}{\tblwidth}{@{\extracolsep{\fill}}lrrrrrrr}
\hline
Scenario & Total demand & \multicolumn{2}{c}{Baseline pressure (bar)} & $\alpha^{*}_{\mathrm{MYNTS}}$ & $\alpha^{*}_{\mathrm{surr}}$ & $p_{\min,\mathrm{MYNTS}}$ & $p_{\min,\mathrm{surr}}$\\
 & (Nm$^3$/s) & $p_{\min}^{\mathrm{base}}$ & $p_{\max}^{\mathrm{base}}$ &  &  & (bar) & (bar)\\
\hline
Max demand & 601.14 & 25.37 & 81.59 & 1.18 & 1.21 & 16.03 & 16.03\\
NW-dominated & 499.19 & 34.90 & 79.88 & 1.26 & 1.25 & 16.08 & 16.01\\
SE-dominated & 553.87 & 39.26 & 75.31 & 1.52 & 1.46 & 16.06 & 16.01\\
NE-dominated & 552.05 & 23.13 & 76.46 & 1.19 & 1.17 & 16.03 & 16.02\\
SW-dominated & 562.67 & 18.71 & 79.32 & 1.03 & 1.02 & 16.08 & 16.01\\
\hline
\end{tabular*}
\end{table}

Table~\ref{tab:loadability} summarizes the results. Across the five stress cases, the surrogate estimates the critical loading factor with absolute errors of 0.01--0.06 (mean 0.026), and the minimum-pressure value at the estimated feasibility boundary differs from MYNTS by at most 0.07~bar. These results indicate that, within the tested one-dimensional demand-scaling paths, the surrogate captures the feasibility crossing with sufficient accuracy for pre-screening; they should not be interpreted as validation of a general operational optimization policy.

\begin{figure}[ht]
\centering
\includegraphics[width=\linewidth]{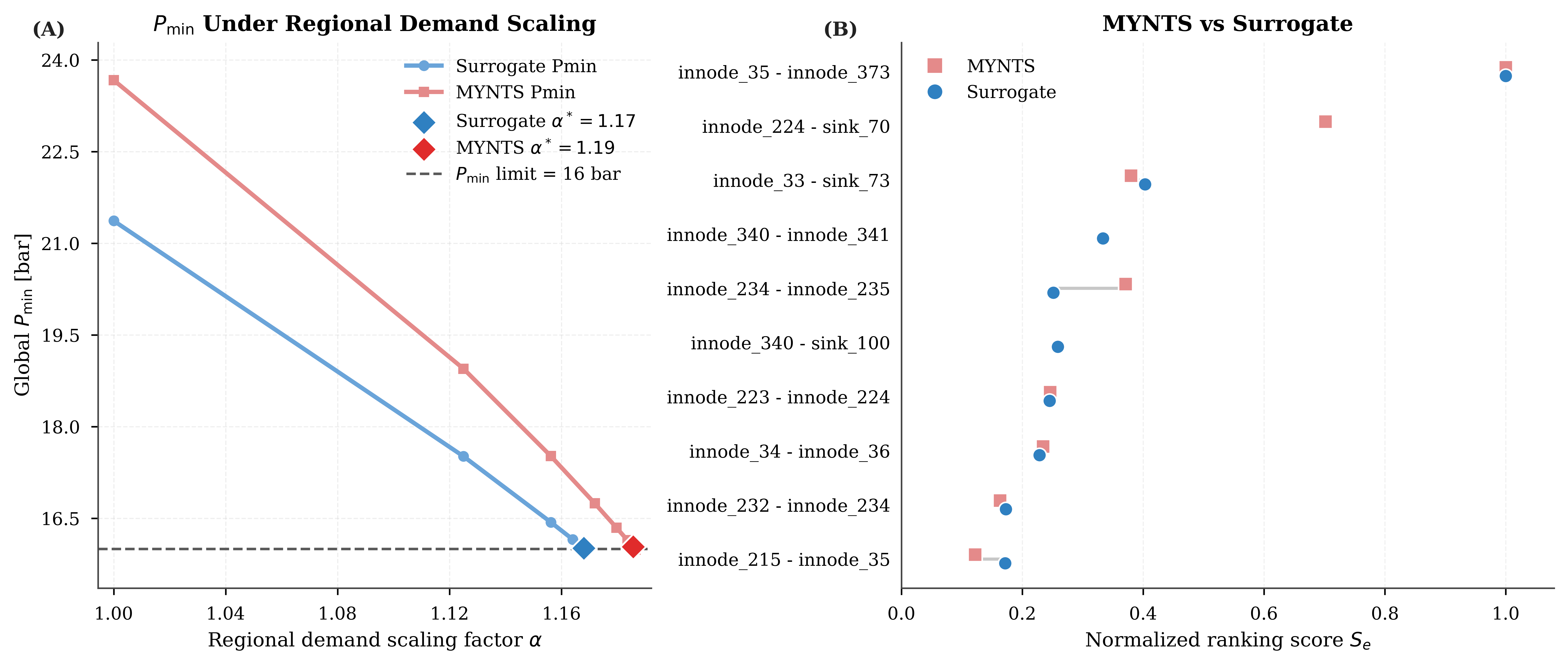}
\caption{Representative feasibility screening and bottleneck ranking for the NE-dominated GasLib-582 scenario. (A) Minimum nodal pressure $P_{\min}$ (bar) versus regional demand scaling factor $\alpha$, comparing MYNTS and surrogate trajectories against the feasibility limit $P_{\min}=16$ bar; diamonds mark the estimated critical loading factors ($\alpha^*_{\mathrm{surr}}=1.17$, $\alpha^*_{\mathrm{MYNTS}}=1.19$). (B) Normalized bottleneck scores for selected pipes, comparing the MYNTS forward stress score with the surrogate sensitivity-based score after both are scaled to $[0,1]$; labels identify the pipe endpoints used in the ranking comparison.}
\label{fig:screening}
\end{figure}

Figure~\ref{fig:screening}(A) illustrates a representative NE-dominated scenario. The surrogate reproduces the monotonic degradation of the minimum nodal pressure with increasing demand and identifies the threshold crossing near the feasibility limit. Although the absolute minimum-pressure trajectory exhibits a visible offset relative to MYNTS, the surrogate captures the feasibility crossing and dominant pressure-decline trend that govern system-level constraints.

Beyond estimating feasibility margins, the surrogate enables identification of limiting network components. A forward bottleneck metric is defined based on the increase in squared-pressure drop relative to baseline conditions,
\begin{equation}
S_e = \left| (\Delta p^2)_e^{\mathrm{limit}} - (\Delta p^2)_e^{\mathrm{base}} \right|,
\end{equation}
which captures the accumulation of hydraulic stress as the system approaches the feasibility boundary.

To complement this forward view, an inverse sensitivity-based ranking is constructed using the differentiable structure of the surrogate. At operating points near the feasibility limit, a smooth objective is defined as the negative soft minimum of nodal pressures within the target region $\mathcal{D}$,
\begin{equation}
J(\mathbf{p}) = -\,\mathrm{softmin}_{i\in\mathcal{D}}\, p_i,
\end{equation}
which serves as a differentiable approximation of the minimum-pressure feasibility constraint. Gradients of this objective with respect to edge-level squared-pressure differences $(\Delta p^2)_e$ are computed via automatic differentiation, and each edge is assigned a sensitivity score,
\begin{equation}
S_{e}^{\mathrm{inv}} = \left|\frac{\partial J}{\partial (\Delta p^2)_e}\right|\,\cdot\,\left|(\Delta p^2)_e\right|,
\end{equation}
which quantifies both the magnitude of local pressure drop and its influence on the critical pressure margin.

Figure~\ref{fig:screening}(B) compares the surrogate inverse-sensitivity ranking with a MYNTS forward stress-based ranking obtained by evaluating hydraulic stress after forward simulations near the feasibility boundary. The surrogate-based inverse ranking captures the dominant bottleneck structure and identifies the same critical transmission corridors. While differences in absolute ranking scores remain, the relative importance of key edges is preserved, indicating that the surrogate provides a meaningful attribution of system-level constraints.

The forward and inverse formulations provide complementary perspectives. The forward metric reflects observed stress accumulation along the feasibility path, while the inverse metric reveals sensitivity pathways through which local perturbations affect global feasibility. Together, they enable both detection and interpretation of bottlenecks within a unified surrogate framework.

\subsection{Out-of-distribution stress-test evaluation}
\label{sec:ood_stress}

To explore the surrogate’s behavior under operating conditions that deviate from the independent and identically distributed (IID) training scenario pool, we conducted a preliminary stress‑test on GasLib‑582. The aim is not to claim broad distributional generalization, but to examine how the physics‑constrained surrogate responds to planning‑relevant demand shifts after a small amount of targeted refinement.

Two OOD scenario families were generated from high-demand operating conditions:
\begin{itemize}
  \item \textbf{High-load set:} scenarios from the top 15\% of total demand in the 1000-scenario base pool, with all demands increased by up to 20\%. Global mass balance is maintained by adjusting the balancing supply accordingly.
  \item \textbf{Regional-concentration set:} also derived from the same high-demand pool, but 50--60\% of total demand is assigned to one of the four geographic regions (NW, NE, SW, SE) used in the feasibility analysis (Section~3.6). This creates more localized hydraulic stress than the IID demand-distribution procedure.
\end{itemize}

The refinement experiment was initialized from the best surrogate weights obtained with the 5000-scenario training configuration. The model was then fine-tuned for 30 epochs using a mixed refinement set consisting of 1000 base scenarios, 250 high-load scenarios, and 250 regional-concentration scenarios. For each OOD family, 500 generated stress-test scenarios were split evenly into refinement-training and held-out test subsets, so the OOD test scenarios were not used during refinement. Evaluation was performed on 500 IID base scenarios, 250 held-out high-load scenarios, and 250 held-out regional-concentration scenarios.

Table~\ref{tab:ood_stress} reports the resulting predictive accuracy, physical-consistency diagnostics, and feasibility-classification behavior. The high-load case shows a moderate degradation in pressure and flow accuracy relative to the IID test set, while retaining strong feasibility-screening performance. The regional-concentration case is more challenging, indicating that localized demand shifts alter the spatial structure of hydraulic stress more strongly than uniform load increases. In both OOD settings, the mass-balance residual remains at numerical precision, showing that the projection layer continues to enforce nodal conservation under distribution shift.

\begin{table}[width=0.98\linewidth,cols=10,pos=h]
\centering
\caption{OOD stress-test performance on GasLib-582 after 30-epoch refinement from the best 5000-scenario surrogate weights.}
\label{tab:ood_stress}
\small
\setlength{\tabcolsep}{2.7pt}
\renewcommand{\arraystretch}{1.05}
\resizebox{\linewidth}{!}{%
\begin{tabular*}{\tblwidth}{@{\extracolsep{\fill}}lrrrrrrrrr}
\hline
Test set & \shortstack{Pressure\\MAE (bar)} & \shortstack{Pressure\\$R^2$} & \shortstack{Flow\\MAE (Nm$^3$/s)} & \shortstack{Flow\\$R^2$} & \shortstack{Mass residual\\(Nm$^3$/s)} & \shortstack{Mono. viol.\\(\%)$^{1}$} & \shortstack{Solver\\infeas.} & \shortstack{Surrogate\\infeas.} & \shortstack{Feas. acc.\\(\%)$^{2}$}\\
\hline
\shortstack[l]{IID\\(500)} & 1.544 & 0.980 & 1.282 & 0.993 & $1.02\times10^{-5}$ & 1.97 & 0 & 0 & 100.0\\
\shortstack[l]{OOD--Highload\\(250)} & 3.470 & 0.775 & 8.770 & 0.834 & $1.79\times10^{-5}$ & 2.83 & 8 & 10 & 99.2\\
\shortstack[l]{OOD--Regional\\(250)} & 4.600 & 0.607 & 8.313 & 0.836 & $1.80\times10^{-5}$ & 2.81 & 79 & 103 & 88.8\\
\hline
\end{tabular*}%
}
\vspace{2pt}
\parbox{0.98\linewidth}{\footnotesize
$^{1}$~Monotonicity violation rate is computed only for edges with $|\Delta p|>1$~bar.\\
$^{2}$~Feasibility accuracy $=(\mathrm{TP}+\mathrm{TN})/N$, where TP and TN denote correctly classified feasible and infeasible cases.
}
\end{table}

From a feasibility-screening perspective, false negatives are the most critical error mode because they cause solver-infeasible cases to be reported as feasible. In the high-load set, no infeasible scenarios were missed. In the regional-concentration set, the false-negative rate is 2.53\%, corresponding to 2 of 79 infeasible scenarios being classified as feasible. This degradation likely reflects flow patterns that differ more substantially from the training distribution and supports solver verification for regionally concentrated cases near the feasibility boundary.

These results suggest that targeted refinement can extend the surrogate to stressed operating regimes, but also clarify the limits of its use. High-load OOD cases remain suitable for pre-screening after refinement, whereas regionally concentrated stress cases should be treated more conservatively and routed to solver-based verification when they approach feasibility limits.

\section{Discussion}

\subsection{Mechanistic interpretation of surrogate behavior across network regimes}

The surrogate's performance is governed by interactions among network topology, physical constraints, and embedded reconstruction operators. These interactions define distinct operating regimes that help explain the observed divergence between pressure and flow predictions.

A primary distinction arises between branched and meshed network structures. In tree-like networks, nodal mass conservation uniquely determines edge flows for a given balanced injection pattern, yielding a single feasible flow solution. The mass-balance projection therefore collapses raw predictions onto this solution manifold and removes most flow uncertainty; in this regime, constraint enforcement dominates the observed flow behavior. In meshed networks, by contrast, cycle-space degrees of freedom create a non-unique mass-conserving flow space. Similar flow-allocation ambiguity has been noted in network flow optimization and graph-based surrogate modeling, where multiple feasible solutions can exist under identical boundary conditions~\cite{AhujaMagnantiOrlin1993NetworkFlows,Rebeschini2019Laplacian,Kerimov2022TransferableGNN}. In this regime, predictive quality depends more directly on learned edge-level representations, with projection acting as a regularizer rather than as the determinant of the solution.

Pressure reconstruction introduces a second mechanism. Nodal pressures are obtained by Laplacian-based integration of predicted edge pressure differences, which couples local errors across the network. Because the reconstructed pressure field is anchored at a reference node, deviations can accumulate along long topological paths, especially in networks with large graph diameter. Such error propagation is well documented in graph-based physical modeling and message-passing architectures, where long-range dependencies require careful treatment to avoid accuracy degradation~\cite{Wu2021GNNsurvey,Karniadakis2021PIML}. Consequently, pressure prediction exhibits a global sensitivity that is less directly exposed in projected flow predictions, which are stabilized by projection onto the mass-conserving subspace.

These observations reveal a structural asymmetry: flow is mainly constrained by global conservation, whereas pressure depends on the global consistency of local pressure-drop estimates. The proposed architecture exploits this structure by combining data-driven edge predictions with deterministic operators that enforce feasibility and coherence, yielding a hybrid model that balances flexibility with physical fidelity.

\subsection{Accuracy--consistency trade-offs in physics-informed learning}

The results confirm that predictive accuracy and physical consistency are intrinsically coupled in surrogate modeling of physical systems. Unconstrained regression can yield pointwise accurate predictions that nonetheless violate fundamental physical laws, whereas strict constraint enforcement reduces the admissible solution space and may shift the solution away from the unconstrained regression optimum. This trade-off is widely recognized in physics-informed machine learning and constrained learning frameworks~\cite{Willard2022ACMSurvey,RaissiPerdikarisKarniadakis2019PINNs}.

Mass-balance projection is central to shaping this trade-off. By confining flows to the feasible subspace, it eliminates large-scale inconsistencies and enforces a coherent network-wide structure. However, projection alone does not guarantee compatibility with the pressure field, which explains the loss of pressure accuracy and the rise in monotonicity violations observed in partially constrained variants. Similar error shifting toward unconstrained variables has been reported in constraint-aware neural networks~\cite{Mohan2023HardConstraints,LuPestourie2021HardPINN}.

The full model mitigates this tension by combining mass-balance projection with Laplacian-based pressure reconstruction and pressure--flow regularization. This coupling enforces complementary structure on flows and pressures, delivering balanced accuracy, feasibility, and physical coherence. The outcome stems from coordinated enforcement of multiple constraints rather than from optimizing any single metric.

From an application standpoint, this asymmetry is critical because conservation violations can invalidate downstream decisions even when pointwise regression metrics appear acceptable. Surrogate evaluation for gas networks must therefore combine predictive error with physical diagnostics; accuracy alone is insufficient.

\subsection{Practical scope for planning workflows}

The planning results represent a controlled demonstration of how a physically constrained surrogate can support repeated hydraulic assessment, rather than a complete replacement for validated simulation tools. The loadability and bottleneck studies show that the model preserves the dominant pressure-margin and stress-ranking patterns along the tested demand-scaling paths, while the OOD stress tests indicate that targeted refinement can extend this screening role to selected high-demand regimes. The surrogate is therefore best suited for early-stage screening, sensitivity analysis, and prioritization of scenarios that warrant full hydraulic verification, especially when localized stress patterns or feasibility-boundary cases are encountered.

The same design logic could support hybrid workflows involving active components such as compressors, regulators, or valves. For example, compressor-station set points, regulator targets, or valve statuses could be introduced as scenario-conditioning variables, while dedicated component models would represent the corresponding control actions and hydraulic constraints. This would make the surrogate useful for providing candidate operating states or informative initializations in subsequent equation-based simulation and optimization, including tasks such as compressor-station set-point optimization. Such extensions would require explicit active-component representations and validation under relevant control policies. The present results therefore establish a basis for surrogate-assisted planning analysis, while deployment-level issues involving control interactions, rare operating regimes, and solver coupling remain for future work.

\subsection{Limitations and future research directions}

The present study is restricted to steady-state, isothermal gas flow with fixed component configurations. Active elements such as compressors, regulators, and valves are not explicitly modeled or controlled during training, which limits the surrogate's direct applicability to systems where pressure regulation and flow control play a central role. In addition, all training and evaluation are based on synthetic benchmark scenarios rather than measured operational data. These choices make the experiments reproducible and help isolate the surrogate mechanism, but they also leave open questions about robustness under operational noise, control actions, broader contingency conditions, and rare operating regimes beyond the targeted high-load and regional-concentration stress tests considered here.

Extending the framework beyond this setting requires both richer physical modeling and more diverse validation data. Transient applications---for example, linepack dynamics, compressor dispatch, valve switching, or contingency analysis---introduce additional state variables and temporal dependencies, requiring coupled spatial--temporal modeling. Non-isothermal effects should also be incorporated for applications where temperature affects density, pressure losses, or operating envelopes. This is particularly relevant for hydrogen transport, hydrogen--natural-gas blends, and \(\mathrm{CO}_2\) networks, where thermodynamic properties and operating constraints can differ substantially from conventional natural gas systems.

Several architectural limitations remain. Although mass conservation is enforced exactly, pressure--flow monotonicity and related pipe-level physical properties are imposed only through soft loss terms, so small residual violations can persist when the corresponding losses are not sufficiently dominant. Future work should therefore investigate stronger constraint-enforcement strategies, including projection-based correction layers, constrained loss formulations, or lightweight feasibility checks coupled to the surrogate. Pressure prediction also depends on global Laplacian reconstruction, which makes the reconstructed field sensitive to accumulated local errors in networks with large graph diameter. Multi-scale reconstruction, local correction mechanisms, or hybrid formulations that combine local pressure-gradient prediction with coarse global pressure information could improve robustness.

Topology generalization remains another open question. The present experiments train and test on fixed benchmark topologies, so the model may learn topology-specific correlations in addition to transferable hydraulic relationships. A stronger validation setting would train across diverse, varying network topologies and evaluate on withheld networks with different loop structures, injection patterns, and pressure-boundary placements. Demonstrating this capability would strengthen the case for graph-based surrogates as reusable planning tools rather than one-off network-specific accelerators.

The effective interaction range of standard message-passing layers may also become a bottleneck in very large transmission networks. Hydraulic effects can propagate over long paths, whereas finite-depth message passing exchanges information locally; simply increasing depth can introduce over-smoothing, over-squashing, and optimization difficulty. Future architecture development should therefore examine graph-attention mechanisms, including GAT and GATv2, or transformer-style graph layers that adaptively weight neighbor contributions and better capture long-range dependencies~\citep{velickovic2018graph,brody2022attentive}. These mechanisms could be combined with physics-informed input features, such as distance to pressure-controlled nodes, distance to major supply or demand regions, or approximate hydraulic resistance along relevant paths. For very large systems, regional decomposition with coupling variables may also help combine local hydraulic detail with system-level interactions.

At the deployment level, the surrogate has so far been validated only for steady-state planning workflows, primarily early-stage screening and sensitivity analysis. Its suitability for operational settings involving active controls, discrete switching, or real-time set-point optimization has not yet been established. Deployment would require explicit compressor and valve models, tested control policies, comparison against independent simulation tools, and calibration or validation against measured system data. Broader rare operating regimes and contingency scenarios also remain insufficiently tested beyond the targeted OOD stress cases considered in this study. Future work should therefore bridge planning-oriented surrogate analysis with operational deployment by addressing solver coupling, control interactions, and data-driven calibration using real network measurements.

\section{Conclusion}

This study developed a physics-informed graph neural network surrogate for steady-state gas-network simulation and planning-oriented feasibility assessment. The framework addresses a key limitation of unconstrained data-driven surrogates by embedding physical reconstruction operators directly into the prediction pipeline, rather than relying on regression accuracy alone.

The proposed model uses an edge-centric representation aligned with gas-flow physics: it predicts pipe-level squared-pressure differences and flows, reconstructs nodal pressures through a Laplacian solve, and projects pipe flows onto the nodal mass-balance constraint. This hybrid formulation enforces exact mass conservation and topologically consistent pressure fields, while pressure--flow compatibility is promoted through physics-guided regularization. The result is a surrogate that combines learned flexibility with deterministic feasibility constraints.

Evaluation on GasLib-134, GasLib-135, and GasLib-582 shows that the model achieves accurate hydraulic predictions while preserving physical consistency. On GasLib-582 with 5000 training scenarios, the model reaches a pressure MAE of 1.05~bar ($R^2=0.981$), corresponding to approximately 1.3\% of the realized dataset pressure range, and a projected-flow $R^2$ of 0.972. The mass-balance projection reduces nodal residuals to numerical precision, on the order of \(10^{-5}\)--\(10^{-4}\)~Nm$^3$/s. Runtime is reduced from seconds for MYNTS to milliseconds per scenario, with the largest benchmark evaluated in less than 40~ms.

The planning demonstrations further show that the surrogate can reproduce feasibility boundaries and identify dominant bottleneck structures along regional demand-scaling paths. The OOD stress-test results indicate that targeted refinement can preserve useful screening behavior under selected high-load conditions, while regionally concentrated demand shifts remain more challenging and should trigger solver-based verification near feasibility limits. These findings support use of the surrogate for high-volume scenario screening, sensitivity analysis, and prioritization of cases that require full hydraulic verification. The surrogate should therefore be viewed as a physically grounded planning accelerator and companion to validated solvers, rather than as a complete replacement for detailed operational simulation.

Several limitations define the next stage of development. The present work considers steady-state, isothermal conditions on fixed benchmark topologies with synthetic operating scenarios, and active components such as compressors, regulators, and valves are not yet explicitly modeled. Future research should extend the framework toward stronger constraint enforcement, topology generalization, long-range graph architectures, active-component modeling, transient and non-isothermal dynamics, and validation against measured operational data. Addressing these issues will be essential for moving from planning-oriented surrogate analysis toward broader deployment in gas, hydrogen, and carbon dioxide infrastructure studies.

\section*{Data and code availability}
The code and reproducibility package supporting this study are archived on Zenodo at DOI: \url{https://doi.org/10.5281/zenodo.20075820}. The development version of the repository is available at: \url{https://github.com/cceekkigg/PIGNN-Gas-Surrogate}.

\section*{Acknowledgements}
The authors gratefully acknowledge the use of computational resources provided by the Fraunhofer Cluster CINES for conducting the simulations presented in this work.

\bibliographystyle{unsrt}
\bibliography{references}

\clearpage
\printcredits
\clearpage

\appendix
\setcounter{table}{0}
\renewcommand{\thetable}{A.\arabic{table}}
\section*{Appendix A. Reproducibility summary}

The full implementation is provided in the accompanying public GitHub repository: \url{https://github.com/cceekkigg/PIGNN-Gas-Surrogate}. This appendix summarizes only the key implementation details required to interpret and reproduce the results reported in the manuscript. Full command-line instructions, configuration files, software dependencies, and reproducibility scripts are maintained in the repository.

\subsection*{A.1 Scenario generation and evaluation setup}

\begingroup
\footnotesize
\setlength{\tabcolsep}{4pt}
\renewcommand{\arraystretch}{0.95}
\captionsetup{type=table}
\begin{center}
\caption{Scenario-generation, data-split, and evaluation setup used in the reported experiments.}
\label{tab:appendix_data}
\begin{tabularx}{\textwidth}{@{}>{\raggedright\arraybackslash}p{0.24\textwidth} >{\raggedright\arraybackslash}X >{\raggedright\arraybackslash}p{0.24\textwidth}@{}}
\hline
Item & Setting & Notes \\
\hline
Benchmark networks &
GasLib-134, GasLib-135, GasLib-582 &
  \\

Reference-pressure setpoints &
GasLib-134: 25--60~bar; GasLib-135 and GasLib-582: 20--80~bar &
Sampled at the pressure-controlled reference node \\

Scenario construction &
Stochastic demand scaling and distribution; balancing supply adjusted to satisfy global mass balance &
Only feasible MYNTS solutions retained \\

Scenario filtering &
Pressure-span filter removes nearly uniform pressure profiles &
Ensures informative hydraulic variation \\

Training sizes &
1000, 2000, and 5000 feasible scenarios per network &
Each split into 80\% training and 20\% validation \\

Test set &
500 independent feasible scenarios per network &
 \\

Repeated runs &
Three independent random seeds  &
24, 33, 42 \\

Loadability screening &
Five high-stress GasLib-582 scenarios; \(p_{\min}=16\)~bar &
 \\
\hline
\end{tabularx}
\end{center}
\endgroup

\subsection*{A.2 Model and training configuration}

\begingroup
\footnotesize
\setlength{\tabcolsep}{4pt}
\renewcommand{\arraystretch}{0.95}
\captionsetup{type=table}
\begin{center}
\caption{Model architecture and training configuration.}
\label{tab:appendix_training}
\begin{tabularx}{\textwidth}{@{}>{\raggedright\arraybackslash}X >{\raggedright\arraybackslash}p{0.23\textwidth} >{\raggedright\arraybackslash}X >{\raggedright\arraybackslash}p{0.23\textwidth}@{}}
\hline
\multicolumn{2}{c}{Model} & \multicolumn{2}{c}{Training} \\
\hline
Parameter & Value & Parameter & Value \\
\hline
GNN layer & GINEConv & Optimizer & Adam \\
Message-passing layers & 6 & Learning rate & \(3\times10^{-4}\) \\
Hidden dimension & 256 & Scheduler & ReduceLROnPlateau \\
Aggregation & Additive aggregation & Batch size & 32 \\
Activation & ReLU & Epochs & 50 \\
Dropout & 0.05 & Early stopping & Validation-loss based \\
Pressure reconstruction & Laplacian/pressure-drop integration & Gradient clipping & Norm 1.0 \\
Flow reconstruction & Mass-balance projection & Active loss terms & Pressure-drop, pressure, raw/projected flow, monotonicity \\
Optional penalties & Turbulent scaling and final-state pipe consistency inactive & Loss weights & \(\lambda_{\Delta p^2}=1.5\), \(\lambda_{p^2}=6.0\), \(\lambda_q=4.0\), \(\lambda_{\mathrm{mono}}=1.0\) \\
Pressure/flow scales & \(c_{p^2}\), \(q_0\) from training data & Flow-loss mix & \(\alpha=0.5\) \\
\hline
\end{tabularx}
\end{center}
\endgroup

\subsection*{A.3 Evaluation metrics}

\begingroup
\footnotesize
\setlength{\tabcolsep}{4pt}
\renewcommand{\arraystretch}{0.95}
\captionsetup{type=table}
\begin{center}
\caption{Evaluation metrics and diagnostic definitions.}
\label{tab:appendix_metrics}
\begin{tabularx}{\textwidth}{@{}>{\raggedright\arraybackslash}p{0.18\textwidth} >{\raggedright\arraybackslash}X >{\raggedright\arraybackslash}p{0.22\textwidth}@{}}
\hline
Metric & Definition & Notes \\
\hline
Pressure MAE &
\(\frac{1}{N}\sum_i \left|\sqrt{\max(\hat{p}_i^2,0)}-\sqrt{\max(p_i^2,0)}\right|\) &
Reported in bar \\

Flow MAE &
\(\frac{1}{M}\sum_e |\hat{q}_e-q_e|\) &
Canonical physical edges \\

\(R^2\) &
\(1-\frac{\sum_j(y_j-\hat{y}_j)^2}{\sum_j(y_j-\bar{y})^2+10^{-8}}\) &
Pressure and flow \\

Mass residual &
\(\|A_{\mathrm{phys}}\hat{q}_{\mathrm{phys}}-d\|_2\) &
All physical nodes; the reference equation is omitted only in the projection solve \\

Monotonicity violation &
Fraction of edges with \(\hat{q}_{u\to v}(\hat{p}_u^2-\hat{p}_v^2)<-\tau_{\mathrm{mono}}\) &
Canonical physical edges; squared-pressure convention \\

Runtime &
Forward-pass wall time per graph &
Batch-averaged inference time \\
\hline
\end{tabularx}
\end{center}
\endgroup

\end{document}